\documentclass{ectj}
\usepackage{etex}
\usepackage{breakcites}
\usepackage{tikz}
\usetikzlibrary{arrows,positioning}
\usepackage{bibunits}
\usepackage{chapterbib}

\usepackage{fix-cm}

\usepackage{todonotes}
\usepackage[compatibility=false]{caption}

\usepackage{setspace}
\usepackage{amsfonts}
\usepackage{amsmath}
\usepackage{amssymb}
\usepackage{bm}
\usepackage{bbm}
\usepackage{color}
	\definecolor{MyDarkBlue}{rgb}{0.1,0.2,.65}
\usepackage{hyperref}
	\hypersetup{linkcolor=black, citecolor=MyDarkBlue, urlcolor=black, colorlinks=true}
\usepackage{mathtools}
\usepackage{graphicx}

\usepackage{subfig}
\usepackage{rotfloat}

\usepackage{enumerate}
\usepackage[vmargin={1.5in, 1.5in},hmargin={1.5in, 1.5in}]{geometry}

\usepackage[T1]{fontenc}

\usepackage{lmodern} \normalfont
\DeclareFontShape{T1}{lmr}{bx}{sc} { <-> ssub * cmr/bx/sc }{}

\usepackage{pst-func}
\usepackage{pst-solides3d}

\numberwithin{equation}{section}

\newcommand{\R}{\mathbb{R}}
\newcommand{\cdist}{\overset{d}{\rightarrow}}
\newcommand{\cprob}{\overset{p}{\rightarrow}}

\newcommand{\prob}{\mathbb{P}}
\newcommand{\expec}{\mathbb{E}}

\newcommand{\eye}[1]{\mathbb{I}_{#1}}

\newcommand{\kron}{\otimes}
\newcommand{\vect}{\text{vec}}
\newcommand{\vech}{\text{vech}}

\usepackage{amsfonts,amssymb,graphics,epsfig,verbatim,bm,latexsym,amsmath,url,amsbsy}

\newtheorem{theorem}{Theorem}
\newtheorem{assumption}{Assumption}

\newtheorem{lemma}{Lemma}

\newtheorem{definition}{Definition}

\renewcommand{\thesection}{\arabic{section}}
\renewcommand{\theequation}{\arabic{section}.\arabic{equation}}

\received{February ???}
\accepted{November ???}
\volume{20}

\title{Projection Inference for Set-Identified SVARs. }

\author[Gafarov, Meier and Montiel Olea]{Bulat~Gafarov$^{\dagger}$, Matthias~Meier$^{\ddagger}$ and Jose~Luis~Montiel~Olea$^{\S}$}

\address{$^{\dagger}$University of California, Davis, Department of Agricultural and Resource Economics.}
\email{bgafarov@ucdavis.edu}

\address{$^{\ddagger}$University of Mannheim, Department of Economics.}
\email{m.meier@uni-mannheim.de}

\address{$^{\S}$Cornell University, Department of Economics.}
\email{montiel.olea@gmail.com}

\def\AmSTeX{$\cal A$\kern-.1667em\lower.5ex\hbox{$\cal M$}\kern-.125em
            $\cal S$-\TeX}
\def\BibTeX{{\rm B\kern-.05em{\sc i\kern-.025em b}\kern-.08em
            T\kern-.1667em\lower.7ex\hbox{E}\kern-.125emX}}

\definecolor{RefColor}{RGB}{35,35,35}
\definecolor{RespColor}{RGB}{0,76,153}

\begin{document}

\begin{abstract}

We study the properties of the classical \emph{projection} method to conduct simultaneous inference about the coefficients of the structural impulse-response function and their  identified set in Structural Vector Autoregressions. We show that---as the sample size grows large---projection inference produces regions for the structural parameters and their identified set with both frequentist coverage and  robust  Bayesian credibility of at least $1-\alpha$. 
We then  calibrate  the radius of the Wald ellipsoid to guarantee that---for a given posterior on the reduced-form parameters---the robust Bayesian credibility of the projection method is exactly $1-\alpha$. We illustrate the main results of the paper using a demand/supply model of the U.S.~labor market.
    
\end{abstract}

\section{{Introduction}}

A Structural Vector Autoregression (SVAR) (\cite{Sims:1980}) is a time series model that brings theoretical restrictions into a linear, multivariate autoregression. The theoretical restrictions are used to transform the \emph{reduced-form} parameters of the multivariate autoregression (regression coefficients and the covariance matrix of residuals) into \emph{structural} parameters that are more amenable to policy interpretation. Depending on the restrictions imposed, the map between reduced-form and structural parameters can be one-to-one (a point-identified SVAR) or one-to-many (a set-identified SVAR). 

It is now customary for empirical macroeconomic studies to impose sign and/or equality restrictions on structural dynamic responses in SVARs in order to set-identify the model, as in the pioneering work of \cite{Faust:1998} and \cite{uhlig:2005}. The vast majority of these studies use numerical methods to construct Bayesian posterior credible sets for the coefficients of the structural impulse-response function. 

Despite the popularity of the Bayesian approach, a practical concern is the fact that posterior inference for the structural parameters continues to be influenced by prior beliefs even if the sample size is infinite. This point has been documented---in detail and generality---in the work of \cite{giacomini_kitagawa:2014}.\footnote{See also \cite{Poirier:1998}, \cite{Gustafson:2009}, and \cite{Moon_Schorfheide:2012}}   \cite{Baumeister_Hamilton:2014} also provide an explicit characterization of the influence of prior beliefs on posterior distributions for structural parameters in set-identified SVARs.   

This paper studies the properties of the classical and well-known \emph{projection method} to conduct \emph{simultaneous} inference about the coefficients of the structural impulse-response function (and their  identified set). The projection method does not rely on the specification of prior beliefs for set-identified parameters. The concrete proposal is to `project' a typical Wald ellipsoid for the reduced-form parameters of a VAR. As we explain below, the suggested nominal $1-\alpha$ projection region consists of all the structural parameters of interest compatible with the reduced-form parameters in a nominal $1-\alpha$ Wald ellipsoid. 

Our main result shows that a nominal $1-\alpha$ projection region has---asymptotically and under mild assumptions---both frequentist coverage and \emph{robust} Bayesian credibility of at least $1-\alpha$. Moreover, building on   \cite{Kaido_Molinari_Stoye:2014}, we show that our baseline projection can be `calibrated\textquoteright\:to eliminate excessive robust Bayesian credibility. 

The remainder of the paper is organized as follows. Section \ref{section:Literature} presents an overview of the projection approach. Section \ref{section:Model} presents the SVAR model and establishes the frequentist coverage of projection. Section \ref{section:BC} establishes the asymptotic robust Bayesian credibility of the projection region.  Section \ref{section:Calibration} presents the `calibration\textquoteright\:algorithm designed to eliminate the excess of robust Bayesian credibility. Section \ref{section:Implementation} applies projection in the context of the demand/supply SVAR for the U.S.~labor market. Section \ref{section:conclusion} concludes.

 \section{{Overview and Related Literature}} \label{section:Literature}

\subsection{{Overview}}

Let $\mu$ denote the parameters of a reduced-form vector autoregression; i.e., the slope coefficients in the regression model and the covariance matrix of residuals. Let $\lambda$ denote the structural parameter of interest; i.e., the response of some variable $i$ to a structural shock $j$ at horizon $k$ (or a vector of responses). In set-identified SVARs there is a known map between  $\mu$ and the lower and upper bounds for the components of $\lambda$; see \cite{uhlig:2005}. Consequently, the smallest and largest value of a particular structural coefficient of interest can be written, simply and succinctly, as $\underline{v}(\mu)$ and $\overline{v}(\mu)$.

Our projection region for $\lambda$ (and for its identified set) is based on a straightforward application of the classical idea of \emph{projection} inference; see \cite{scheffe1953method}, \cite{dufour1990exact}, and \cite{Dufour:2005, Dufuour:2007}. Let $\widehat{\mu}_{T}$ denote the sample least squares estimator for $\mu$ and let CS$_{T}(1-\alpha; \mu)$ denote its nominal $1-\alpha$ Wald confidence ellipsoid. If, asymptotically, CS$_{T}(1-\alpha; \mu)$ covers the parameter $\mu$ with probability $1-\alpha$, then, asymptotically, the interval
\begin{equation} \label{equation:ProjectionCS}
\text{CS}_{T}(1-\alpha; \lambda) \equiv \Big[ \inf_{\mu \in \text{CS}_{T}(1-\alpha, \mu)} \underline{v}(\mu)  \: , \: \sup_{\mu \in \text{CS}_{T}(1-\alpha, \mu)} \overline{v}(\mu) \Big]
\end{equation}
covers the set-identified parameter $\lambda$ (and its identified set) with probability of at least $1-\alpha$ (uniformly over a large class of data generating processes).\footnote{The application of projection inference to SVARs was first suggested by \cite{Moon_Schorfheide:2012} (p. 11, NBER working paper 14882). The projection approach is also briefly mentioned in the work \cite{Kline_Tamer:2015} (Remark 8) in the context of set-identified models. None of these papers established the properties for projection inference discussed in our work.}  

In many applications there is interest in conducting \emph{simultaneous} inference on $h$ structural parameters; for example, if one wants to analyze the response of variable $i$ to a structural shock $j$ for all horizons ranging from period 1 to $h$  as in \cite{Jorda:2009}, \cite{Kilian_Inoue:2013, Inoue:2014}, and \cite{Lutkepohl:2015}. In this case, one can show that the projection region is given by: 
\begin{equation}\label{equation:ProjectionCSjoint}
\text{CS}_{T}(1-\alpha; (\lambda_1, \ldots, \lambda_h)) \equiv \text{CS}_{T}(1-\alpha; \lambda_1) \times \ldots \times \text{CS}_{T}(1-\alpha; \lambda_h),
\end{equation} 
which covers the structural coefficients $(\lambda_1, \ldots \lambda_h )$ and their identified set with probability at least equal to $1-\alpha$ as the sample size grows large. The only assumption required to guarantee the frequentist coverage of our projection region is the asymptotic validity of the confidence set for the reduced-form parameters, $\mu$.

{\scshape General Applicability:} The validity of our projection method requires no regularity assumptions (like continuity or differentiability) on the bounds of the identified set $\underline{v}(\cdot)$ and $\overline{v}(\cdot)$. This means we can handle the typical application of set-identified SVARs in the empirical macroeconomics literature (exclusion restrictions on contemporaneous coefficients, long-run restrictions, elasticity bounds, and of course sign/zero restrictions on the responses of different variables at different horizons for different shocks).

{\scshape Computational Feasibility:} The implementation of our projection approach requires neither numerical inversion of hypothesis tests nor sampling from the space of rotation matrices. Instead, we use state-of-the-art optimization algorithms to solve for the maximum and minimum value of a mathematical program to compute the two end points of the confidence interval in ($\ref{equation:ProjectionCS}$). 

{\scshape Robust Bayesian Credibility:} In the spirit of making our results appealing to Bayesian decision makers---and following the seminal work of \cite{giacomini_kitagawa:2014}---we show that our suggested nominal $1-\alpha$ projection region will have---as the sample size grows large---robust Bayesian credibility of at least $1-\alpha$. This means that the asymptotic posterior probability that the vector of structural parameters of interest belongs to the projection region will be at least $1-\alpha$; for a fixed prior on the reduced-form parameters, $\mu$, and for \emph{any prior} on the set-identified parameters. A sufficient condition to establish the robust Bayesian credibility of projection is that the prior for $\mu$ used to compute credibility satisfies the \emph{Bernstein-von Mises} theorem.

{\scshape `Calibrated\textquoteright\:Projection:} Despite the features highlighted above, projection inference is \emph{conservative} both for a frequentist and a robust Bayesian. Both the asymptotic confidence level and the asymptotic robust credibility of projection can be strictly above $1-\alpha$. \cite{Kaido_Molinari_Stoye:2014} [henceforth, KMS] refer to the excess frequentist coverage as \emph{projection conservatism} and develop an innovative \emph{calibration} approach to eliminate it.\footnote{Another paper proposing a procedure to eliminate the frequentist excess coverage in moment-inequality models is \cite{Bugni_Canay_Shi:2014}. Adapting their \emph{profiling} idea to our set-up could be of theoretical interest and of practical relevance. We leave this question open for future research.  } 
The calibration exercise in KMS requires, in the SVAR context, the computation of Monte-Carlo coverage probabilities for the projection region over an exhaustive grid of values for the reduced-form parameters, $\mu$. In several SVAR applications, the dimension of $\mu$ compromises the construction of an exhaustive grid. 
Instead of insisting on removing excessive frequentist coverage, we suggest practitioners to calibrate projection to attain a robust Bayesian credibility of exactly $1-\alpha$. Broadly speaking, the calibration consists of drawing $\mu$ from its posterior distribution (or a suitable large-sample Gaussian approximation); evaluating the functions $\underline{v}(\mu), \overline{v}(\mu)$ for each draw of $\mu$; and decreasing the radius defining the projection region until it contains exactly $(1-\alpha)\%$ of the values of $\underline{v}(\mu), \overline{v}(\mu)$ (for different horizons and different shocks if desired). 

{\scshape Illustrative Example:} The illustrative example in this paper is a simple demand and supply model of the U.S.~labor market. We estimate standard Bayesian credible sets for the dynamic responses of wages and employment using the Normal-Wishart-Haar prior specification in \cite{uhlig:2005} and also the alternative prior specification recently proposed by \cite{Baumeister_Hamilton:2014}. The main set-identifying assumptions are sign restrictions on contemporaneous responses: an expansionary structural demand shock increases wages and employment upon impact; an expansionary structural supply shock decreases wages but increases employment, also upon impact.\footnote{Following \cite{Baumeister_Hamilton:2014} we further consider bounds on the wage elasticity of both labor demand and labor supply, and bounds on the long-run impact of a demand shock on employment. } 

The Bayesian credible sets for this application illustrate the attractiveness of set-identified SVARs. The data, combined with prior beliefs, and with the (set)-identifying assumptions imply that the initial responses to demand and supply shocks persist in the medium-run, which was not restricted ex-ante. 
The Bayesian credible sets for this application also illustrate how the quantitative results in set-identified SVARs can be affected by the prior specification. For example, under the prior in \cite{Baumeister_Hamilton:2014} the 5-year ahead response of employment to a demand shock could be as large as $4\%$; whereas under the priors in \cite{uhlig:2005} the same effect is at most $2\%$.

Our baseline projection approach allows us to get a prior-free assessment about the magnitude (and direction) of the  structural responses of interest. For example, the largest value in our projection region for the 5-year response of employment to a demand shock is around 2.5\%. This effect is larger than the one implied by the prior in \cite{uhlig:2005}, but smaller than the one implied by the priors in \cite{Baumeister_Hamilton:2014}.

Our baseline projection approach---though informative about the effects of demand shocks---is not conclusive about the medium-run effects of structural supply shocks on wages and employment (the projection region allows for both positive and negative responses). This could be a consequence of either the robustness of projection or its conservativeness. To disentangle these effects, we calibrate projection to guarantee that it has exact robust Bayesian credibility. The calibrated projection shows that an expansionary supply shock will decrease wages in each quarter over a 5 year horizon. However, the qualitative effects of supply shocks on employment remain undetermined. The simple SVAR for the labor market illustrates the usefulness of both the baseline and the calibrated projection to analyze the robustness of quantitative and qualitative results in SVARs to prior beliefs. 

 \subsection{{Related Literature}} 
There continues to be interest in departing from the standard Bayesian analysis of set-identified SVARs in an attempt to provide robustness to the choice of priors. Below we provide a short description of the similarities and differences between our projection approach and three alternative methods available in the literature.

a) In a pioneering paper, \cite{Moon-Granziera-Schorfheide:2013} [MSG] proposed \label{R1.2} a Bonferroni frequentist inference using a moment-inequality, minimum distance framework based on \cite{Andrews_Soares:2010}.\footnote{An earlier working paper version of MSG also considered a projection of the \cite{Andrews_Soares:2010} statistic for moment inequalities for each hypothetical value of the structural matrix $B$. Instead, our paper projects Wald ellipsoid for reduced form parameters $\mu$. Our approach has a computational advantage, since we avoid
computationally costly grid search step over values of $B$.} In terms of applicability, their procedures are designed for set-identified SVARs that impose restrictions on the dynamic responses of only one structural shock. It is possible to extend their approach to the same class of models that we consider; there is, however, a serious issue regarding computational feasibility. Specifically, the Bonferroni approach in MSG requires the researcher to compute---by simulation---a critical value for each single orthogonal matrix of dimension $n \times n$, where $n$ is the dimension of the SVAR. Our baseline implementation of the projection method does not require any type of grid over the space of orthogonal matrices and does not require the simulation of any critical value.

b) The seminal paper of \cite{giacomini_kitagawa:2014} [GK] develops a novel and generally applicable robust Bayesian approach to conduct inference about a specific coefficient of the impulse-response function in a set-identified SVAR. In terms of our notation, their procedure can be described as follows. One takes posterior draws from $\mu$ and evaluates, at each posterior draw, the functions $\underline{v}(\mu), \overline{v}(\mu)$ by solving a nonlinear program. Their credible set is a numerical (grid-search) approximation to the \emph{smallest} interval that covers $100(1-\alpha) \%$ of the posterior realizations of the identified set.   

Our baseline procedure will be typically faster to implement than the GK robust procedure (since our baseline projection only needs to solve two nonlinear programs). The price to pay for the reduced computational cost is the excess of robust Bayesian credibility.  
Our calibrated projection requires a similar amount of work as the GK robust method (both procedures evaluate the bounds of the identified set for each posterior draw). Our main contribution relative to \cite{giacomini_kitagawa:2014} is that our calibrated projection allows for \label{R1.5} \emph{non-conservative simultaneous} credibility statements over different horizons, different variables, and different shocks.

c) \cite{GMM:2015} [GMM1] establish the differentiability of the bounds of $\underline{v}(\mu), \overline{v}(\mu)$ for a class of SVAR models that impose restrictions only on the responses to one structural shock. Based on the differentiability results, they propose a `delta-method' confidence interval given by the plug-in estimators of the bounds plus \label{R1.12c} or minus standard errors multiplied by the normal critical value $r$. It can be shown that, in large samples, the `delta-method' procedure in GMM1 is equivalent to a projection region based on a Wald ellipsoid for $\mu$ with radius $r^2$.\footnote{See details in Appendix C of \cite{gafarov2025projectioninferencesetidentifiedsvars}.}

\section{{Basic Model, Main Assumptions, and Frequentist Results}} 
\label{section:Model}

\subsection{{Model}} \label{subsection:Model}

This paper studies the $n$-dimensional Structural Vector Autoregression with $p$ lags; i.i.d.~structural innovations---denoted $\varepsilon_{t}$---distributed according to $F$; and unknown $n \times n$ structural matrix $B$:
\begin{equation}\label{equation:SVAR}
Y_t= A_1 Y_{t-1} + \ldots + A_{p} Y_{t-p} +   B \varepsilon_{t}, \: \: \expec_{F}[\varepsilon_{t}]=0_{n \times 1}, \: \: \expec_{F}[\varepsilon_t \varepsilon_t' ] \equiv  \eye{n}, 
\end{equation}

\noindent see \cite{Lutkepohl:2007}, p. 362.

The \emph{reduced-form parameters} of the SVAR model are defined as the vectorized autoregressive coefficients and the half vectorized covariance matrix of reduced-form residuals:
$$\mu \equiv ( \text{vec}(A)', \text{vech}(\Sigma)' )^{\prime} \in \R^{d}, \quad \text{where} \quad A \equiv (A_1, A_2, \ldots , A_p), \quad \Sigma \equiv BB'.$$
These parameters can be estimated directly from the data using least squares: 
$$\widehat{\mu}_{T} \equiv   ( \text{vec}(\widehat{A}_{T})', \text{vech}(\widehat{\Sigma}_{T})' )^{\prime},$$ 
\noindent where
\begin{eqnarray*}
\widehat{A}_{T} &\equiv& \Big(  \frac{1}{T} \sum_{t=1}^{T} Y_t X_t' \Big) \Big( \frac{1}{T} \sum_{t=1}^{T} X_t X_t' \Big)^{-1},  \quad \widehat{\Sigma}_{T} \equiv \frac{1}{T} \sum_{t=1}^{T} \widehat{\eta}_t \widehat{\eta}_t', 
\end{eqnarray*}
with $X_t  \equiv  (Y_{t-1}', \ldots , Y_{t-p}')'$ and $\widehat{\eta}_t  \equiv Y_t - \widehat{A}_{T} X_t$.

A common formula for the asymptotic variance of $\widehat{\mu}_{T}$ in stationary models is:
$$\widehat{\Omega}_{T} \equiv V_{T} \Big( \frac{1}{T} \sum_{t=1}^{T} \text{vec} \Big( [\widehat{\eta}_t X_t',  \widehat{\eta}_t\widehat{\eta}_t' - \widehat{\Sigma}_{T} ] \Big) \text{vec} \Big( [\widehat{\eta}_t X_t',  \widehat{\eta}_t\widehat{\eta}_t' - \widehat{\Sigma}_{T} ] \Big)' V_{T}'$$
where 
$$ V_T \equiv \begin{pmatrix} \Big(\frac{1}{T} \sum_{t=1}^{T} X_t X_t' \Big)^{-1}   \kron \eye{n}& \textbf{0} \\ \textbf{0} & L_n  \end{pmatrix},$$
and $L_n$ is the matrix of dimension $n(n+1)/2 \times n^2$ such that $\vech(\Sigma)=L_n \vect(\Sigma)$, see \cite{Lutkepohl:2007}, p. 662 equation A.12.1. 

\subsection{{Assumptions for frequentist inference}}

The SVAR parameters $(A_1, \ldots , A_p, B, F)$ define a probability measure, denoted $P$, over the data observed by the econometrician. The measure $P$ is assumed to belong to some class $\mathcal{P}$ which we describe in this section.\\

We state a simple high-level assumption concerning the asymptotic behavior of the $1-\alpha$ Wald confidence ellipsoid for $\mu$, which is defined as:
\begin{equation} \label{equation:Wald}
CS_{T}(1-\alpha; \mu) \equiv \Big\{ \mu \in \R^{d} \quad | \quad T (\widehat{\mu}_{T}-\mu)' \widehat{\Omega}_{T}^{-1} (\widehat{\mu}_{T}-\mu) \leq \chi^2_{d,1-\alpha}   \Big\}.\footnotemark
\end{equation}
\footnotetext{The radius $\chi^2_{d,1-\alpha}$ denotes the $1-\alpha$ quantile of a central $\chi^2$ distribution with $d$ degrees of freedom.}  
The first assumption requires \emph{uniform consistency in level} (over the class $\mathcal{P}$) of the Wald confidence set for the reduced-form parameters. That is:

\begin{assumption} \label{ass:A1} $ \liminf_{T \rightarrow \infty} \inf_{P \in \mathcal{P}} P \Big( \mu(P) \in CS_{T}(1-\alpha; \mu) \Big) \geq 1-\alpha.$
\end{assumption}

Assumption \ref{ass:A1} holds if the class $\mathcal{P}$ under consideration contains only \emph{uniformly} stable VARs where the error distributions, $F$, have  \emph{uniformly} bounded fourth moments.\footnote{A class $\mathcal{P}$ that satisfies Assumption \ref{ass:A1} could be written by using a uniform version of the conditions in \cite{Lutkepohl:2007}, p. 73. This is, there are positive constants $c_1, c_2, c_3, c_4$ such that:
\begin{eqnarray*}
\mathcal{P} &=& \{ (A_1, A_2, \ldots , A_p, B, F) \: | \: \quad  \text{det}(\eye{n}- A_1z-\ldots A_p z^{p}) \notin (-c_1, c_1) \text{ for } z \in \mathbb{C}, |z| \leq 1; \\
&& B \text{ is such that } 0<c_2<\text{eigmin}(BB^{\prime})<\text{eigmax}(BB^{\prime})<c_3; \text{and } \expec_{F}[|\varepsilon_{n_1,t} \varepsilon_{n_2,t} \varepsilon_{n_3,t} \varepsilon_{n_4,t}|]<c_4\\
&& \text{ for all } t \text{ and } n_1, n_2, n_3, n_4 \in \{1, \ldots n\}, \text{ and } \expec_{F}[\varepsilon_t]=\bf{0}_{n \times 1}, \: \expec_{F}[\varepsilon_{t} \varepsilon_{t}^{\prime}] = \eye{n} \quad \}.
\end{eqnarray*}
Other possible definitions of $\mathcal{P}$ can be given by generalizing Theorem 3.5 in \cite{Chen2015} to either multivariate linear processes with i.i.d. innovations or to martingale difference sequences. For joint inference on  VAR parameters $\mu$ in non-stationary case see \cite{gafarov2024wildinferencewildsvars}.} Assumption \ref{ass:A1} turns out to be sufficient to conduct frequentist inference on the structural parameters of a set-identified SVAR, defined as follows. \\

{\scshape Coefficients of the Structural Impulse-Response Function:} Given the autoregressive coefficients $A \equiv (A_1, A_2, \ldots , A_p)$ define, recursively, the nonlinear transformation
$$C_k(A) \equiv \sum_{m=1}^{k} C_{k-m}(A) \: A_{m}, \quad k \in \mathbb{N},$$
\noindent where $C_0= \eye{n}$ and $A_{m}=0$ if $m>p$; see \cite{Lut90IRF}, p. 116.

\begin{definition} [{Coefficients of the Structural IRF}] The $(k,i,j)$-\emph{coefficient of the structural impulse-response function} is defined as the scalar parameter:
$$\lambda_{k, i, j}(A,B) \equiv  e'_{i}C_{k}(A) Be_{j},$$ 
where $e_i$ and $e_j$ denote the $i$-th and $j$-th column of the identity matrix $\eye{n}$. 
\end{definition}

\subsection{{Main result concerning frequentist inference}}

In this section we show that, under Assumption \ref{ass:A1}, it is possible to `project' the $1-\alpha$ Wald confidence set for $\mu$ to conduct frequentist inference about the coefficients of the structural impulse-response function and the function itself in set-identified models.  

{\scshape Set-Identified SVARs:} As mentioned in the introduction, the SVAR allows researchers to transform the reduced-form parameters, $\mu \equiv (\vect(A)',\vech(\Sigma)^{\prime})^{\prime}$, into the structural parameters of interest, $\lambda_{k,i,j}(A,B)$. The parameter $\mu$ determines a unique value of $A$; however, several values of $B$ are compatible with $\Sigma$ (any $B$ such that $BB'=\Sigma$). This indeterminacy of $B$ implies there are multiple values of $\lambda_{k,i,j}(A,B)$ that are compatible with one value of $\mu$.

{\scshape{The Identified Set and its Bounds:}} It is common in applied macroeconomic work to impose restrictions on the matrix $B \in \R^{n \times n}$ in order to limit the range of a structural coefficient of interest, $\lambda_{k,i,j}$ (taking $\mu$ as given). Mathematically, a set of restrictions on $B$---that we denote as $\mathcal{R}(\mu)$---can be interpreted as a subset of $\R^{n \times n}$. This leads to the following definition:

\begin{definition} [{Identified Set and its bounds}] Fix a vector of reduced-form parameters, $\mu$, and a set of restrictions $\mathcal{R}(\mu)$ on $B$.
\begin{enumerate}[a)]
\item The \emph{identified set} for the structural parameter $\lambda_{k,i,j}(A,B)$ is defined as:
\begin{equation}
\mathcal{I}_{k,i,j}^{\mathcal{R}} (\mu) \equiv \Big\{ v \in \R \: \Big | \: v=\lambda_{k,i,j}(A,B), \: \: BB'=\Sigma, \: \: \text{and } B \in \mathcal{R} (\mu) \Big\}.
\end{equation}

\item The \emph{upper bound} of the identified set $\overline{v}_{k,i,j}(\mu)$ is defined as the value function of the program:
\begin{equation}\label{equation:PointEstimate}
\overline{v}_{k,i,j}(\mu) \equiv \sup_{B \in \R^{n \times n}} e_i'C_k(A) B e_j, \quad \text{s.t}. \quad BB'=\Sigma, \: \text{and } B \in \mathcal{R} (\mu).
\end{equation}
The \emph{lower bound}, $\underline{v}_{k,i,j}(\mu),$ is defined analogously. \\

\item Consider any collection $\lambda^H \equiv \{\lambda_{k_h,i_h,j_h}\}_{h=1}^{H}$ of structural coefficients and let its identified set be given by:
$$ \mathcal{I}_{H}^{\mathcal{R}} (\mu) \equiv \Big\{ (v_1, \ldots , v_{H}) \in \R^{H} \: \Big | \: v_h=\lambda_{k_h,i_h,j_h}(A,B), \: BB'=\Sigma, \: \: \text{and } B \in \mathcal{R} (\mu) \Big\}.  $$ 

\end{enumerate}
\end{definition}

\begin{table}[b] 
\caption{{\scshape Common Restrictions Used in Set-Identified SVARs}}\label{table:T1}
\small
\vspace{-4mm}
\begin{center}
\begin{tabular}{lll} 
\hline\hline \vspace{-2mm}\\
{\scshape Restrictions} & \textbf{Notation} & \textbf{Examples}   \\

Short-run & $e_i'Be_j = \: 0$ \text{ or } $e_i'{B'}^{-1}e_j = \: 0$ & \cite{Sims:1980}, \cite{Arias:2015},\\  & &   \cite{Christiano-Eichenbaum-Evans:1996} 
\\ \vspace{-2mm} \\
Long-run & $e_i'(\eye{n}-A_1-\ldots A_p )^{-1}Be_j= 0$ & \cite{BQ:1989} \\
\\
Sign & $e_i'C_k(A) Be_j \geq ,\leq 0$ & \cite{uhlig:2005}, \\ & & \cite{Mountford_Uhlig:2009} \\
\vspace{-2mm} \\
Elasticity Bounds & $\frac{e_i'C_k(A) Be_j }{e_{\tilde{i}}'C_k(A) Be_j}  \geq, \leq c, \: \: \tilde{i} \neq i$ & \cite{Kilian_Murphy:2012}, \\ & & \label{R1.8b} \cite{Baumeister2024}  
\\ \vspace{-2mm} \\
Shape Constraints & e.g., $e_i'C_k(A)Be_j \leq e_i'C_{k+1}(A)Be_j$ & \cite{Scholl:08} \\
\vspace{-0mm} \\
Other & $e_i'(\eye{n}-A_1-\ldots A_p )^{-1}Be_j \geq, \leq 0$ &  \\
 & $e_i'C_k(A) Be_j = 0$ &  \\
 & $g(B,\mu) \geq , \leq , = 0$ &  \\
\vspace{-2mm}\\ \hline\hline \vspace{-6mm}\\  
  \end{tabular}
\end{center}\textbf{Note:} {\footnotesize Subscript $i$ denotes the variable, $j$ denotes the shock, and $k$ the horizon. The projection approach can handle SVAR models with any of the restrictions described in this table (imposed on one or multiple shocks).}
\end{table}

Table \ref{table:T1} presents a list of the most common restrictions, $\mathcal{R}(\mu)$, used in SVAR analysis (all of which can be handled by our frequentist approach described below).\footnote{\label{R1.8d}Our approach can further handle restrictions on the FEVD as in \cite{Volpicella2022} and ranking restrictions of \cite{amir2021identification}. One can also extend our approach to alternative normalization assumptions provided the identified set is bounded, see \cite{read2024set}.}
\label{R1.1a}Throughout the paper, we focus on the case with non-mutually contradictory identification restrictions under probability measure $P$. 
In other words, the structural restrictions are correctly specified, and the identified set is non-empty.
It is not critical for the validity of the inference procedure, but it guarantees non-empty confidence sets in sufficiently large samples. 
If the identified set is truly empty, the projection confidence set could also be empty, but it remains valid since an empty set is always a subset of any set.
An empty confidence set would imply model misspecification.
Practitioners could omit some of the constraints to restore non-emptiness.\footnote{In the case of misspecified moment inequalities, \cite{10.1093/restud/rdad033} provides a valid  procedure for projection inference on pseudo-parameters. Their procedure can also be adapted to our setup for inference on pseudo-parameters. } 
\label{R2.1}
Note that under Assumption \ref{ass:A1}, we allow for sequences of drifting DGP $P_T$ and the corresponding $\mu_T$ that correspond to non-singleton sets $\mathcal{I}_{H}^{\mathcal{R}} (\mu_T)$, but with a singleton limit as considered in \citep{gafarov2014identification}. 
One can also accommodate both strong and weak proxy-IV variables in our setup by explicitly including them in the VAR system and imposing appropriate short run restrictions.

{\scshape Projection Approach:} A key feature of set-identified SVARs is that the bounds of the identified set depend on a finite-dimensional parameter. `Projecting\textquoteright\:down the $1-\alpha$ Wald ellipsoid for $\mu$ seems a natural approach to conduct inference on the structural impulse response function. The first result in this paper establishes the frequentist uniform validity of projection inference. 

\begin{theorem} [{Frequentist Coverage of Projection Inference for $\lambda^{H}$}] \label{thm:freqCover} \mbox{}\\*
Consider the projection region for the collection of structural coefficients $\lambda^H \equiv \{\lambda_{k_h,i_h,j_h}\}_{h=1}^{H}$ given by:
\begin{equation}\label{equation:ProjectionCSJoint}
CS_{T}(1-\alpha, \lambda^{H} ) \equiv CS_{T}( 1-\alpha, \lambda_{k_1,i_1,j_1} ) \times \ldots \times CS_{T}( 1-\alpha, \lambda_{k_H,i_H,j_H} ) \subseteq \R^{H},
\end{equation}
where
\begin{equation} \label{equation:ProjectionCSResult}
\text{CS}_{T}(1-\alpha; \lambda_{k,i,j}) \equiv \Big[ \inf_{\mu \in \text{CS}_{T}(1-\alpha, \mu)} \underline{v}_{k,i,j}(\mu)  \: , \: \sup_{\mu \in \text{CS}_{T}(1-\alpha, \mu)} \overline{v}_{k,i,j}(\mu) \Big],
\end{equation}
and $CS_{T}(1-\alpha; \mu)$ is the $1-\alpha$ Wald confidence ellipsoid for $\mu$. If the class of data generating processes $\mathcal{P}$ satisfies  Assumption \ref{ass:A1}, then:
$$\liminf_{T \rightarrow \infty} \inf_{P \in \mathcal{P}} \inf_{ \lambda^{H} \in \mathcal{I}^{R}_{H}(\mu(P))}  P \Big( \lambda^{H} \in CS_{T}(1-\alpha; \lambda^{H}) \Big) \geq 1-\alpha.$$
That is, the projected confidence interval in (\ref{equation:ProjectionCSJoint}) covers the vector of structural coefficients $\lambda^H$ with probability at least $1-\alpha$, uniformly over the class $\mathcal{P}$. 
\end{theorem}

\begin{proof}
The proof of Theorem \ref{thm:freqCover} uses a standard and conceptually straightforward projection argument. Take an element $P \in \mathcal{P}$ and let $\lambda^{H} \in \R^{H}$ be any given element of the identified set $\mathcal{I}_{H}^{\mathcal{R}} (\mu(P))$. Note that:
$$P \Big( \lambda^H \in CS_{T}(1-\alpha; \lambda^{H}) \Big)$$
\begin{eqnarray*}
&=& P \Big( (\lambda_{k_1, i_1, j_1} , \ldots, \lambda_{k_H, i_H, j_H})  \in CS_{T}(1-\alpha; \lambda_{k_1,i_1, j_1}) \times \ldots \times CS_{T}(1-\alpha; \lambda_{k_H,i_H, j_H})   \Big) \\
&& \Big( \text{by definition of our confidence interval for $\lambda^{H}$} \Big) \\
& \geq & P \Big( [\underline{v}_{k_h,i_h,j_h}(\mu(P)) \: , \:  \overline{v}_{k_h,i_h,j_h}(\mu(P))] \subseteq \Big[ \inf_{\mu \in \text{CS}_{T}(1-\alpha, \mu)} \underline{v}_{k_h,i_h,j_h}(\mu)  \: , \: \\ 
&& \hspace{6.2cm} \sup_{\mu \in \text{CS}_{T}(1-\alpha, \mu)} \overline{v}_{k_h,i_h,j_h}(\mu) \Big] \quad \forall h=1, \ldots ,H \Big),\\
&& \Big(\text{since $ \lambda_{k_h,i_h,j_h} \in [\underline{v}_{k_h,i_h, j_h}(\mu(P)), \overline{v}_{k_h,i_h, j_h}(\mu(P))]$} \Big) \\
&\geq & P \Big( \mu(P) \in CS_{T}(1-\alpha; \mu)  \Big).
\end{eqnarray*}

\noindent The desired result follows directly from Assumption \ref{ass:A1}. This shows that the projection region for $\lambda^H$ is uniformly consistent in level.  
\end{proof}

The idea of `projecting' a confidence set for a parameter $\mu$ to conduct inference about a lower dimensional parameter $\lambda$ has been used extensively in econometrics; see \cite{scheffe1953method}, \cite{dufour1990exact}, and \cite{Dufour:2005, Dufuour:2007} for some examples. In addition to its conceptual simplicity, one advantage of the projection approach is that its validity does not require special conditions on the identifying restrictions that can be imposed by practitioners. For instance, we do not need to assume that $\underline{v}_{k,i,j}(\cdot)$ and $\overline{v}_{k,i,j}(\cdot)$ are continuous or differentiable functions of the reduced-form parameters.

The problem of conducting inference on the whole impulse-response function (and not only on one specific  coefficient) has been a topic of recent interest, both from the Bayesian and frequentist perspective, as exemplified below. 
For Bayesian set-identified SVARs with only sign restrictions, \cite{Kilian_Inoue:2013} report the vector of structural impulse-response coefficients with highest posterior density (based on a prior on reduced-form parameters and a uniform prior on rotation matrices). They propose a Bayesian credible set (represented by \emph{shotgun} plots) that characterizes the joint uncertainty about a given collection of structural impulse-response coefficients. 

For frequentist point-identified SVARs, \cite{Inoue:2014} propose a bootstrap procedure that allows the construction of asymptotically valid confidence regions for any subset of structural impulse responses. To the best of our knowledge, our projection approach is the first frequentist procedure for set-identified SVARs that provides confidence regions for any collection of structural coefficients (response of different variables, to different shocks, over different horizons).

It is important to note that \cite{uhlig:2005}'s approach to conduct inference on set-identified SVARs does not provide credible sets for vectors of the structural parameters. The same is true for the Bayesian approaches described in the recent work of \cite{Arias:2015} and \cite{Baumeister_Hamilton:2014}, as well as the approaches of \cite{Moon-Granziera-Schorfheide:2013} and \cite{giacomini_kitagawa:2014}.

    A common concern in set-identified models is whether the suggested inference approach is valid only for the identified parameter, $\lambda^H$, or also for its identified set $\mathcal{I}_{H}^{\mathcal{R}} (\mu)$. Note that the second-to-last inequality in the proof of Theorem \ref{thm:freqCover} imply that our projection region covers the identified set of any vector of coefficients $\lambda^H$.

\section{ {Robust Bayesian Credibility}}
\label{section:BC}
This section analyzes the \emph{robust credibility} of projection as the sample size grows large. \\

{\scshape Bayesian Set-up:} In a Bayesian SVAR the distribution of the structural innovations is fixed and treated as a known object. A common choice---which we follow in this section---is to assume that $F \sim \mathcal{N}_n(0, \eye{n})$. We discuss how to relax this restriction after stating Assumption \ref{ass:A2}. 

Let $P^*$ denote some prior for the structural parameters $(A_1, \ldots ,A_p, B)$ and let $\lambda^{H}(A,B)$ $\in \R^{H}$ denote the vector of structural coefficients of interest. For a given square root of $\Sigma \equiv B B^{\prime}$ define the `rotation\textquoteright\:matrix $Q \equiv \Sigma^{-1/2} B$. It is well known that a prior $P^*$ can be written as $(P^*_{\mu}, P^*_{Q|\mu})$, where $P^*_{\mu}$ is a prior on the reduced-form parameters, and $P^*_{Q|\mu}$ is a prior on the rotation matrix, conditional on $\mu$.
Following this notation, let $\mathcal{P}(P^*_{\mu})$ denote the class of prior distributions such that $\mu\sim P^*_{\mu}$. 

We are interested in characterizing the smallest posterior probability that the set $\text{CS}_{T}(1-\alpha; \lambda^H)$ could receive, allowing the researcher to vary the prior for $Q$:
\begin{equation} \label{equation:credibility}
\inf_{P^* \in \mathcal{P}(P^*_\mu)}P^* \Big( \lambda^{H}(A,B) \in CS_T (1-\alpha; \lambda^H) \: \Big| \: Y_1, \ldots , Y_T \Big).
\end{equation}  
The event of interest is whether the structural coefficients $\lambda^H (A,B)$ (treated as random variables in the Bayesian Set-up) belong to the projection region, after conditioning on the data. This event would typically be referred to as the credibility of $\text{CS}_{T}(1-\alpha; \lambda^H)$ (see \cite{berger1985statistical}, p. 140). We would like to find the smallest credibility of projection when different priors over $Q$ are considered as in the pioneering work of \cite{Kitagawa:2012}. We follow  the recent work of \cite{giacomini_kitagawa:2014} and refer to (\ref{equation:credibility}) as the \emph{robust Bayesian credibility} of the set $\text{CS}_T (1-\alpha, \lambda^H)$.

Let $f(Y_1, . . . , Y_T | \mu)$ denote the Gaussian statistical model for the data (which depends solely on the reduced-form parameters) and let $o_{p}(1; Y_1, \ldots Y_T | \mu )$ denote a random variable such that $\lim_{T \rightarrow \infty} P_{Y_1, \ldots, Y_{T}|\mu}(|o_{p}(1; Y_1, \ldots Y_T | \mu )|$ $> \epsilon)=0$ for all $\epsilon>0$ when the distribution of the data is conditioned on $\mu$.

{\scshape Main Assumption for Bayesians:} Robust credibility can be viewed as a random variable (as it depends on $Y_1, \ldots, Y_{T}$). We use the following high-level assumption to characterize its asymptotic behavior: 

\begin{assumption} \label{ass:A2} Whenever $Y_1, \ldots, Y_T \sim f( Y_1, \ldots ,Y_{T} | \mu_0) $, the prior $P^*$ is such that:
$$P^*\Big( \mu(A,B) \in CS_T(1-\alpha; \mu) \: \Big| \: Y_1, \ldots , Y_{T}  \Big) = 1-\alpha + o_{p}(1; Y_1, \ldots, Y_T | \mu_0 ).$$ 
\end{assumption}

Assumption \ref{ass:A2} requires the prior over the reduced-form parameters (and the statistical model) to be regular enough to guarantee that the asymptotic Bayesian credibility of the $1-\alpha$ Wald ellipsoid converges in probability to $1-\alpha$. Thus, our high-level assumption is implied by the Bernstein-von Mises Theorem (\cite{Dasgupta08}, p. 291) for the reduced-form parameter $\mu$. 

Since the Gaussian statistical model $f(Y_1, \ldots Y_{T} | \mu_0)$ can be shown to be Locally Asymptotically Normal (LAN) whenever $A_0$ is stable and $\Sigma_0$ has full rank, Theorem 1 and 2 in \cite{ghosal1995convergence} (GGS) imply that Assumption \ref{ass:A2} will be satisfied whenever $P^*_{\mu}$ has a continuous density at $\mu_0$ with polynomial majorants.\footnote{In Online Appendix S1 we verify an `almost sure\textquoteright\:version of Assumption \ref{ass:A2} for a Gaussian SVAR for the Normal-Wishart priors suggested in \cite{Uhlig:1994} and a confidence set for $\mu$ based on the formula for the asymptotic variance $\widehat{\Omega}_{T}$ that obtains in the Gaussian model \citep[][p.~93]{Lutkepohl:2007}. } In fact, the same theorems could be used to establish Assumption \ref{ass:A2} for non-Gaussian SVARs that are LAN and satisfy the regularity conditions of \cite{Ibragimov:2013} (IH), as long as $\text{CS}_{T}(1-\alpha; \mu)$ is centered at the Maximum Likelihood estimator of $\mu$ and $\widehat{\Omega}_{T}$ is replaced by the model\textquoteright s inverse information matrix. An alternative approach to establish Assumption \ref{ass:A2} using a different set of primitive conditions can be found in \cite{Ben:2016}. 

We now establish the robust Bayesian credibility of projection as $T \rightarrow \infty$. 

\begin{theorem}[{Asymptotic Robust Bayesian Credibility of Projection}] \label{result:R3} \mbox{}\\*
Suppose that the prior $P^*$ for $(A,B)$ satisfies Assumption \ref{ass:A2} at $\mu_0$. Then:
$$\inf_{P^* \in \mathcal{P}(P^*_\mu)} P^* \Big( \lambda^{H}(A,B) \in CS_T (1-\alpha; \lambda^H) \: \Big| \: Y_1, \ldots , Y_T \Big) \geq 1-\alpha + o_{p}(1; Y_1, \ldots Y_T | \mu_0 ).$$ 
\end{theorem}

\begin{proof}
Note that:
$$ \inf_{P^* \in \mathcal{P}(P^*_\mu)} P^* \Big( \lambda^H (A,B) \in CS_{T}(1-\alpha; \lambda^{H}) \: \Big |  \: Y_1, \ldots Y_{T} \Big)$$
\begin{eqnarray*}
&=&\inf_{P^* \in \mathcal{P}(P^*_\mu)} P^* \Big( \lambda_{k_h, i_h, j_h} (A,B)   \in CS_{T}(1-\alpha; \lambda_{k_h,i_h, j_h}) \: \forall \: h=1 \ldots , H \: \Big | \: Y_1, \ldots , Y_{T}   \Big) \\
&& \Big( \text{by definition of the projection region for $\lambda^{H}$} \Big) 
\end{eqnarray*}
\begin{eqnarray*}
& \geq & \inf_{P^* \in \mathcal{P}(P^*_\mu)}P^* \Big( [\underline{v}_{k_h,i_h,j_h}(\mu(A,B)) \: , \:  \overline{v}_{k_h,i_h,j_h}(\mu(A,B))] \in CS_{T}(1-\alpha; \lambda_{k_h,i_h, j_h})  \quad \forall \: h=1 \ldots , \\
&& H \: \Big | \: Y_1, \ldots , Y_T \Big),\\
&& \Big(\text{since $ \lambda_{k_h,i_h,j_h}(A,B) \in [\underline{v}_{k_h,i_h, j_h}(\mu(A,B)), \overline{v}_{k_h,i_h, j_h}(\mu(A,B))]$} \: \text{for any } A,B \Big) \\
&\geq & P^* \Big( \mu(A,B) \in CS_{T}(1-\alpha; \mu)  \: \Big | \: Y_1, \ldots , Y_{T} \Big).
\end{eqnarray*}

Assumption \ref{ass:A2} gives the desired result. 
\end{proof}

This means that---given any prior that satisfies Assumption \ref{ass:A2}---our projection region can be interpreted, in large samples, as a robust $1-\alpha$ credible region for the impulse-response function and its coefficients.

\section{{Calibrated Projection for a Robust Bayesian}}
\label{section:Calibration}

The projection approach generates \emph{conservative} regions for both a frequentist and a robust Bayesian. For a frequentist, the large-sample coverage may be strictly above the desired confidence level. For a robust Bayesian, the asymptotic robust credibility of the nominal $1-\alpha$ projection region may be strictly above $1-\alpha$.  

This section applies the approach in \cite{Kaido_Molinari_Stoye:2014} to eliminate the excess of robust Bayesian credibility in a computationally tractable way. We focus on calibrating the robust credibility of our projection region to be exactly equal to $1-\alpha$ (either in a finite sample for a given prior on $\mu$, or in large samples for a large class of priors on $\mu$).

Given a vector $\Lambda^{H}= \{\lambda_{k_h,i_h,j_h}\}_{h=1}^{H}$ of structural coefficients of interest and its  corresponding nominal $1-\alpha$ projection region, the calibration exercise is based on the following result: \begin{theorem}[{Calibration of Robust Credibility}] \label{result:R2} 
Let $P^*_{\mu}$ denote a prior for the reduced-form parameters. Suppose there is a nominal level $1-\alpha^*(Y_1, \ldots ,Y_T)$ such that for every data realization:
$$P^*_{\mu} \left( \times_{h=1}^{H} [\underline{v}_{k_h,i_h,j_h}(\mu), \overline{v}_{k_h,i_h,j_h}(\mu)] \subseteq \text{CS}_{T}(1-\alpha^*(Y_1, \ldots,Y_T), \lambda^{H})  |  Y_1, \ldots, Y_T  \right)$$
equals $1-\alpha$. Then, for every data realization:
$$\inf_{P^* \in \mathcal{P}(P^*_\mu)}P^* \Big( \lambda^{H}(A,B) \in CS_T (1-\alpha^*(Y_1, \ldots, Y_T); \lambda^H) \: \Big| \: Y_1, \ldots , Y_T \Big) = 1-\alpha.$$
\end{theorem}

\begin{proof}
See  Appendix \ref{subsection:R2}. 
\end{proof}

This means that in order to calibrate the robust credibility of projection, it is sufficient to choose $1-\alpha^*(Y_1, \ldots, Y_T)$ to guarantee that exactly $\alpha\%$ of the bounds of the identified set for the different structural coefficients in $\lambda^H$ fall outside the projection region. 

{\scshape Calibration Algorithm:}  The calibration algorithm we propose consists in finding a nominal level $1-\alpha^*(Y_1, \ldots, Y_{T})$ such that the posterior probability of the event: 
$$ [\underline{v}_{k_1,i_1,j_1}(\mu), \overline{v}_{k_1,i_1,j_1}(\mu)] \times \ldots \times [\underline{v}_{k_h,i_h,j_h}(\mu), \overline{v}_{k_h,i_h,j_h}(\mu)] \subseteq \text{CS}_{T}(1-\alpha^*, \lambda^{H})  $$ 
equals $1-\alpha$ under the posterior distribution associated with the prior $P^*_{\mu}$ or under a suitable large-sample approximation for the posterior, such as $\mu | Y_1, \ldots Y_{T} \sim \mathcal{N}_{d}(\widehat{\mu}_{T}, \widehat{\Omega}_{T}/T)$.\footnote{The Gaussian approximation for the posterior will eliminate projection bias asymptotically, provided a Bernstein-von Mises Theorem for $\mu$ holds. We establish this result in Online Appendix S2.  } \\ 

The calibration algorithm is the following:
\begin{enumerate}
\item  Generate $M$ draws from the posterior of the reduced-form parameters. If desired, one could use the large-sample approximation of the posterior given by:
$$\mu_m^* \sim \mathcal{N}_{d} (\widehat{\mu}_{T}, \widehat{\Omega}_{T}/T).$$
\item Let $\lambda^{H}= \{\lambda_{k_h,i_h,j_h}\}_{h=1}^{H}$ denote the structural coefficients of interest. For each $h=1, \ldots H$ and for each $m=1, \ldots M$ evaluate:
$$ [\underline{v}_{k_h, i_h, j_h}(\mu^*_m), \overline{v}_{k_h, i_h, j_h}(\mu^*_m)],$$
as defined in equation (\ref{equation:PointEstimate}). 
We provide Matlab code to evaluate these bounds \label{R1.9}using an SQP/IP algorithm (described in Online Appendix~S3). 
\item Fix an element $\alpha_s$ on the interval $(\alpha,1)$. Set a tolerance level $\eta>0$.
\item For each $m=1, \ldots M$ generate the indicator function $z_m$ that takes the value of $0$ whenever there exists an index $h \in \{1, \ldots H\}$ such that:
$$ [\underline{v}_{k_h, i_h, j_h}(\mu^*_m), \overline{v}_{k_h, i_h, j_h}(\mu^*_m)] \not \subset \text{CS}_{T}(1-\alpha_s, \lambda_{k_h, i_h, j_h}).$$  It is equal to 1 otherwise.
The projection region CS$_T(1-\alpha_s, \lambda_{k_h, i_h, j_h})$ is defined in equation (\ref{equation:ProjectionCSResult}) in Theorem \ref{thm:freqCover} and implemented using an SQP/IP algorithm \label{R1.12d} (described in Online Appendix~S3). 
\item Compute the robust credibility of the nominal $1-\alpha_s$ projection as: 
$$ RC_{T}(\alpha_s)= \frac{1}{M} \sum_{m=1}^{M} z_m.$$
If such quantity is in the interval $[1-\alpha-\eta, 1-\alpha+\eta]$ stop the algorithm. If $\text{RC}_{T}(\alpha_s)$ is strictly above (below) $1-\alpha + \eta$, go back to Step 3 and choose a larger (smaller) value of $\alpha_{s}$.\\ 
\end{enumerate}

It is also possible to show that whenever the bounds of the identified set for each $\lambda_h $ are differentiable, there is a sense in which our calibration algorithm also removes the excess of frequentist coverage \label{R1.11} (see details in \cite{gafarov2025projectioninferencesetidentifiedsvars}). \\

\section{{Implementation of Baseline and Calibrated Projection}}
\label{section:Implementation}

\subsection{{Projection as a mathematical optimization problem}}

This subsection discusses the implementation of the baseline projection region:
$$\text{CS}_{T}(1-\alpha; \lambda_{k,i,j}) \equiv \Big[ \inf_{\mu \in \text{CS}_{T}(1-\alpha, \mu)} \underline{v}_{k,i,j}(\mu)  \: , \: \sup_{\mu \in \text{CS}_{T}(1-\alpha, \mu)} \overline{v}_{k,i,j}(\mu) \Big].$$
We note that both the upper   and lower bounds of this confidence interval can be thought of as solutions to a pair of `nested' optimization problems. 

The first optimization problem---that we refer to as the \emph{inner} optimization---solves for $\overline{v}_{k,i,j}(\mu)$ and $\underline{v}_{k,i,j}(\mu)$. These functions correspond to the largest and smallest value of the structural impulse response $\lambda_{k,i,j}$ given a set of restrictions and a vector of reduced-form parameters $\mu$. 
The second optimization problem---that we refer to as the \emph{outer} optimization problem---solves for the maximum value of $\overline{v}_{k,i,j}(\cdot)$ and the minimum value of $\underline{v}_{k,i,j}(\cdot)$ over the $(1-\alpha)$ Wald confidence ellipsoid, CS$_T(1-\alpha,\mu)$. 

{\scshape Implementation:} Our proposal is to combine the inner and outer problems into a \emph{single} mathematical program that gives the bounds of the projection confidence interval directly. The upper bound can be found by solving:
\begin{eqnarray}\label{equation:JointOptimization}
\sup_{A, \Sigma, B} e_i'C_k(A) B e_j \quad  \text{subject to} \quad BB' = \Sigma , \quad B \in \mathcal{R}(\mu),  \text{ and } \\
T (\widehat{\mu}_{T}- \mu(A,\Sigma))'\widehat{\Omega}_{T}^{-1} (\widehat{\mu}_{T}- \mu(A,\Sigma)) \leq \chi^2_{d, 1-\alpha}. \notag\\ \notag
\end{eqnarray}
\noindent The lower bound of the projection confidence interval can be found analogously. Importantly, the simple reformulation in (\ref{equation:JointOptimization}) allows us to base the implementation of our projection region upon state-of-the-art solution algorithms for optimization problems. 
\label{R2.2}
For most applications, including the one considered in the paper,  restrictions $B \in \mathcal{R}(\mu)$ are smooth functions. 
This allows one to use a simple  SQP/IP algorithm (see details in Online Appendix S3).
For non-smooth constraints $\mathcal{R}(\mu)$, one can use a more computationally demanding global search algorithm (for example, a genetic algorithm) instead.

\subsection{{Implementing baseline projection in an example}}
As an example, we consider the demand-supply SVAR model studied in Section 5 of \cite{Baumeister_Hamilton:2014} [henceforth, BH]. We fit a 6-lag VAR to U.S.~data on growth rates of real labor compensation, $\Delta w_t$, and total employment, $\Delta n_t$, from 1970:Q1 to 2014:Q2.\footnote{Our selection is based on the fact that 6 is the smallest number of lags such that \text{CS}$(68\%;\mu)$ does not contain unstable VAR coefficients and non-invertible reduced-form covariance matrices. $68\%$ confidence sets are frequently used in applied macroeconomic research. The Bayes Information Criteria and the Akaike Information Criteria both select less than six lags.} 

Using our notation, the demand-supply SVAR can be written as:
$$ \begin{pmatrix} \Delta w_t \\ \Delta n_t \end{pmatrix} = A_1 \begin{pmatrix} \Delta w_{t-1} \\ \Delta n_{t-1} \end{pmatrix} + \ldots + A_6 \begin{pmatrix} \Delta w_{t-6} \\ \Delta n_{t-6} \end{pmatrix} + B \begin{pmatrix} \epsilon^d_{t} \\ \epsilon^s_{t} \end{pmatrix},$$
BH set-identify an expansionary demand and supply shock by means of the following sign restrictions:
$$ B \equiv \begin{pmatrix} b_1 & b_3 \\ b_2 & b_4 \end{pmatrix} \quad \text{satisfies} \quad \begin{bmatrix} + & -\\ + & + \end{bmatrix}.$$ 
The sign restrictions state that a demand shock increases both real labor compensation and total employment, while a supply shock lowers wages but raises employment. 

In this model, the short-run wage elasticity of labor supply (identified from a demand shock) is defined as:
$$ \alpha \equiv b_2/b_1$$ 
Likewise, the short-run wage elasticity of labor demand (identified from a supply shock) is defined as: 
$$\beta \equiv b_4/b_3 $$ 
Finally, the long-run impact of a demand shock on employment is given by:
$$ \gamma \equiv e_2'(\eye{n}-\sum_{p=1}^{6} A_p )^{-1}Be_1.$$

BH impose three additional restrictions. The first two are elasticity bounds motivated by the findings of different empirical studies. \cite{Hamermesh:1996}, \cite{Akerlof:07}, \cite{lichter14} provide bounds on the wage elasticity of labor demand. \cite{chetty:11}, \cite{Whalen:2012} provide bounds on the wage elasticity of labor supply. The third and final restriction arises from imposing lower and upper bounds on the long-run impact of a demand shock on employment. 

BH incorporate the restrictions in the form of priors on the structural parameters, but we treat the constraints as additional sign restrictions. Let $t_v$ denote the standard $t$ distribution with $v$ degrees of freedom. Table \ref{table:T2} summarizes the way in which BH incorporate prior information: 
\begin{table}[h] 
\caption{{\scshape Additional Identifying Restrictions}}\label{table:T2}
\vspace{-4mm}
\begin{center}
\begin{tabular}{cccc} 
\hline\hline \vspace{-2mm}\\
{\scshape Restrictions} & \textbf{Motivation}  &  \textbf{BH} & \textbf{This paper}   \\
 
\vspace{-2mm}\\
Bounds on $\alpha$  & Empirical studies   & $\alpha \sim \max\{.6 + .6 t_{3},0\}$    & $.27 \leq \alpha \leq 2$      \\
& report $\alpha \in [.27, 2]$ & &   \\
\vspace{-2mm}\\ 
Bounds on $\beta$  & Empirical studies   & $\beta \sim \min\{-.6 + .6 t_{3},0\}$    & $-2.5 \leq \beta \leq -.15$      \\
&  $\beta \in [-2.5, -.15]$ & &   \\
\vspace{-2mm}\\
Bounds on $\gamma$   & $\gamma=0$ is too strong   & $\gamma \sim \mathcal{N} (0,V)$    & $-2V \leq \gamma \leq 2V$      \\
\vspace{-2mm}\\ \hline\hline \vspace{-2mm}\\
\end{tabular}
\end{center}
\end{table}

Thus, summarizing, our version of the BH model has 10 sign restrictions:
\begin{eqnarray*}
\text{Demand and Supply Shocks} &:& b_1 \geq 0, b_2 \geq 0, -b_3 \geq 0, b_4 \geq 0,\\
\text{Elasticity Bounds} &:& 2b_1 - b_2 \geq 0, b_2 - .27 b_1 \geq 0, \\
&& b_4 + .15 b_3 \geq 0, -2.5 b_3 - b_4 \geq 0, \\
\text{Long-Run} &:& e_2'(\eye{n}-\sum_{p=1}^{6} A_p )^{-1}Be_1 + 2V \geq 0, \\
&-&e_2'(\eye{n}-\sum_{p=1}^{6} A_p )^{-1}Be_1 + 2V \geq 0,\\
\end{eqnarray*}
where the parameter $V$ is allowed to take the values $\{.01, .1, 1\}$ as in p. 1992 of BH. 

\subsection{{Results of the implementation of baseline projection}} \label{sec:SQP}

Using our SQP/IP local solution algorithm, we compute the 68\% projection confidence intervals for the cumulative response of wages and employment to the structural shocks in the model (20 consecutive quarters and setting $V=1$). In addition to the projection region, we compute the 68\% Bayesian credible set following the implementation in both \cite{uhlig:2005} and BH.

Figure \ref{figure:BH} shows the projection region as a solid blue line and the standard Bayesian credible set (based on BH priors) as a gray-shaded area.  
Online Appendix S4 Figure 3 
shows the boundaries of the projection region as a solid blue line and the Bayesian credible set based on \cite{uhlig:2005}'s priors as a gray-shaded area.  
The 68\% credible sets differ substantially depending on the specification of prior beliefs. Such sensitivity is the main motivation for our projection approach. In this example, the length of the credible sets for the cumulative response of employment seems to differ by a factor of at least two. The projection region seems quite large compared to the credible sets. This could be a consequence of either the robustness of projection or its conservativeness. To disentangle these effects, we calibrate projection to guarantee that it has exact robust Bayesian credibility in the next subsection.

\begin{figure}  
\centering
\caption{68\% Projection Region and 68\% Credible Set.}
\caption*{(\cite{Baumeister_Hamilton:2014} priors)}
{\includegraphics[keepaspectratio, scale=.45]{./figures/figImplementation/IRF_BH15_v1_s1_p6.eps}} \hspace{.5cm}
{\includegraphics[keepaspectratio, scale=.45]{./figures/figImplementation/IRF_BH15_v1_s2_p6.eps}}\\
\subfloat[Expansionary Demand Shock]{\includegraphics[keepaspectratio, scale=.45]{figures/figImplementation/IRF_BH15_v2_s1_p6.eps}} \hspace{.5cm}
\subfloat[Expansionary Supply Shock]{\includegraphics[keepaspectratio, scale=.45]{./figures/figImplementation/IRF_BH15_v2_s2_p6.eps}}\\
\label{figure:BH}
\caption*{\footnotesize ({\scshape Solid, Blue Line}) 68\% Projection Region;  ({\scshape Shaded, Gray Area}) 68\% Bayesian Credible Set based on the priors in \cite{Baumeister_Hamilton:2014}. }
\end{figure} 
  

%

We investigate the computational feasibility of our projection by comparing its computing time with standard Bayesian methods.\footnote{To get a fair sense of the computational cost, none of the global algorithms were parallelized.} Since the global methods are initialized at the local solution, these procedures take at least as much time as SQP/IP.  Among the three global methods considered, the Genetic Algorithm takes the longest. Brute-force grid search (which refers to grid search on CS$_T(1-\alpha, \mu)$ to optimize $\underline{v}_{k,i,j}(\mu)$ and $\overline{v}_{k,i,j}(\mu)$)  with only 1,000 draws from $\mu \in \mathcal{\R}^{27}$ takes about 6 times longer than the baseline SQP/IP and generates substantially smaller bounds. We further compare the accuracy across a range of local and global solution methods. For this application, it seems that none of the global algorithms improve on the local solution obtained from SQP/IP. For details, see Online Appendix S3 Table 1 and Figure 2.

\subsection{{Implementing calibrated projection in our example}}
The key restriction used to set-identify an expansionary demand shock in the illustrative example is that it must increase wages and employment upon impact. According to the credible sets in Figures \ref{figure:BH} and Online  Appendix S4 Figure 3, the demand shock has---in fact---noncontemporaneous effects on these two variables (every quarter over a 5 year horizon). Our calibrated projection confirms that there are medium-run effects of demand shocks on employment but suggests that the non-zero effects on wages beyond the first two quarters could be an artifact of prior beliefs.

\begin{figure} 
\centering
\caption{68\% Projection Region and 68\% Calibrated Projection.}
{\includegraphics[keepaspectratio, scale=.45]{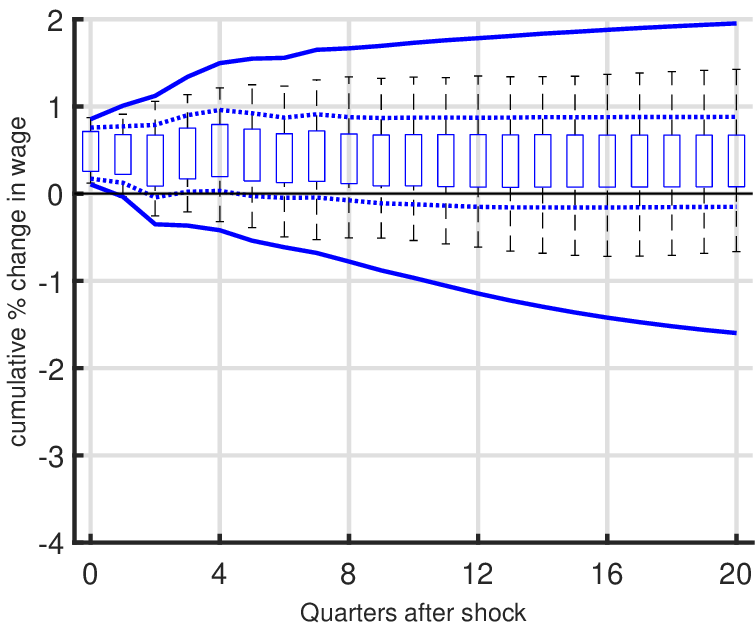}} \hspace{.5cm}
{\includegraphics[keepaspectratio, scale=.45]{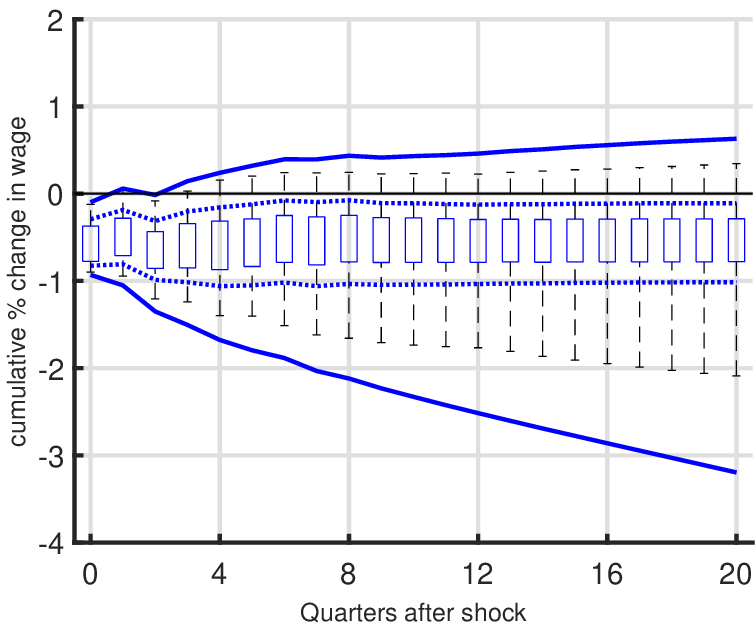}}\\
\subfloat[Expansionary Demand Shock]{\includegraphics[keepaspectratio, scale=.45]{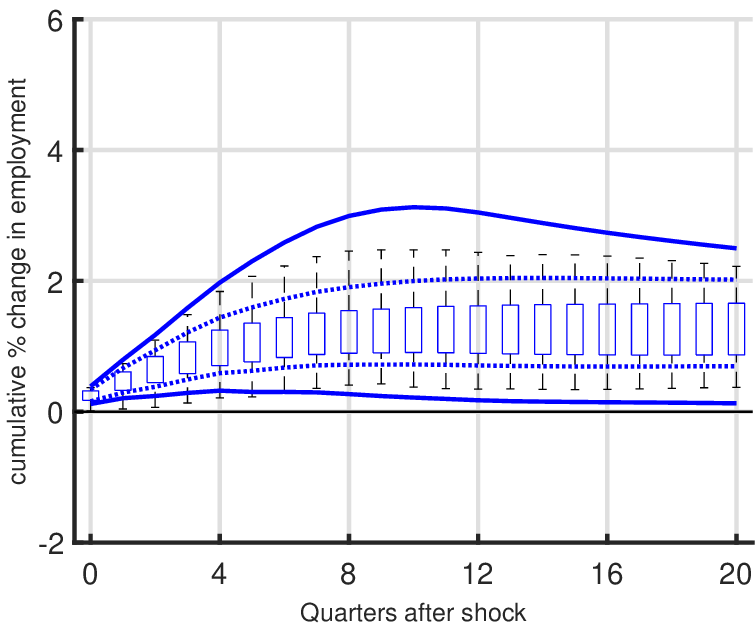}} \hspace{.5cm}
\subfloat[Expansionary Supply Shock]{\includegraphics[keepaspectratio, scale=.45]{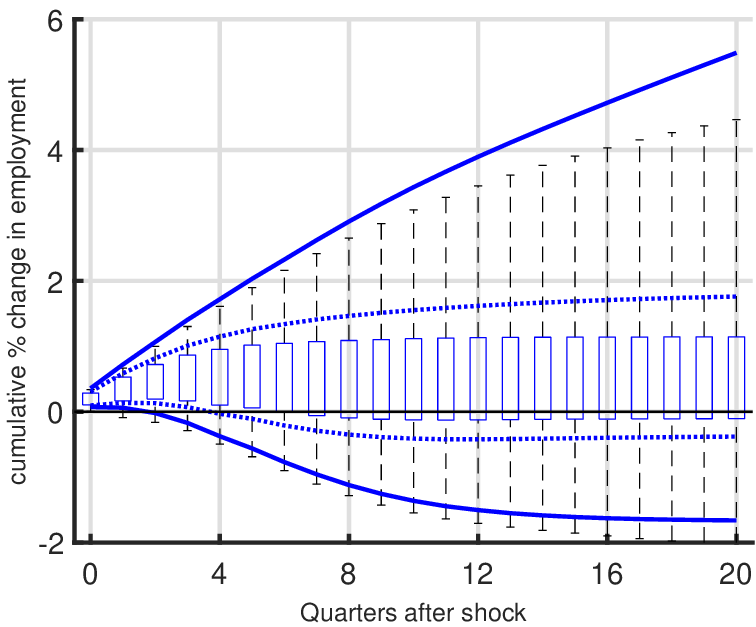}}\\
\label{figure:Calibration}
\caption*{\footnotesize ({\scshape Solid Line}) 68\% Projection region; ({\scshape Dotted Line}) 68\% Projection region \textit{calibrated} to guarantee 68\% robust Bayesian credibility of the IRF functions jointly (100,000 draws from the Gaussian approximation to the posterior of $\mu$); ({\scshape Box}) 68\% Projection region \textit{calibrated} horizon-by-horizon and shock-by-shock; ({\scshape Black Dashed Line}) Support of the bounds of the identified set given the 100,000 posterior draws.}
\end{figure} 

A similar observation is true for supply shocks. Our calibrated projection suggests that the decrease in wages five years after an expansionary supply shock is robust to the choice of prior on the set-identified parameters. The medium-run effects of supply shocks on employment lack this robustness.\

{\scshape Implementation of our Calibrated Projection:} We close this subsection providing further details about the computational demands of our calibration exercise. 

Instead of working with a specific posterior for $\mu$, we calibrated the projection relying on the large-sample approximation $\mu | Y_1, \ldots ,Y_{T} \sim \mathcal{N}_{d}(\widehat{\mu}_{T}, \widehat{\Omega}_{T}/T)$. Taking draws from this model is straightforward and does not require any special sampling technique (as a  Markov Chain Monte-Carlo). Figure \ref{figure:Calibration} uses M=100,000 draws. 

As described in our calibration algorithm, for each of the draws of $\mu$ (denoted $\mu^*_m$), and for each horizon $k\in\{0,1,2, \ldots 20\}$, variable $i \in \{\text{wage},\text{employment}\}$ and shock $j \in \{\text{demand shock},\text{supply shock}\}$ we solved two mathematical programs to generate:
$$ [\underline{v}_{k, i, j}(\mu^*_m), \overline{v}_{k, i, j}(\mu^*_m)].$$
Computing the bounds of the identified set for all the combinations $(k,i,j)$ given $\mu^*_m$ took approximately 9 seconds. Generating the boxes and the black dashed lines in Figure \ref{figure:Calibration} \label{R1.3e}took approximately 5 hours using 50 parallel Matlab `workers' on a computer cluster.\footnote{Calibrating projection to guarantee frequentist coverage at one point in the parameter space took us 76 hours using the 50 parallel Matlab  workers in the same computer cluster.} Notice that we chose M=100,000 for illustrative purposes, and the calibration results are barely different for M=1,000, which takes 3 minutes using the same computer cluster (or 2.5 hours without parallelization at all).

After generating the bounds of the identified set, the calibration exercise adjusts the nominal level of projection to simultaneously contain $68\%$ of the draws from the bounds of the identified set for each combination $(k,i,j)$.\footnote{To do this, we ran the baseline projection SQP/IP algorithm for different nominal confidence levels. An efficient calibration algorithm that requires only a few iterations over the nominal level is the combination of bisection with secant and interpolation, as provided by Matlab's \texttt{fzero} function. For a reasonably low tolerance of $\eta=0.001$, we need 15 iteration steps. With each step taking about 734 seconds (see Online Appendix Table 1), steps 3 through 5 take about 1 hour.} The calibrated confidence level for the Wald ellipsoid is $1.85 \cdot 10^{-4}$\% instead of the original 68\%. 



\section{{Conclusion}} \label{section:conclusion} 

A practical concern regarding standard Bayesian inference for set-identified Structural Vector Autoregressions is the fact that prior beliefs continue to influence posterior inference even when the sample size is infinite. Motivated by this observation, this paper studied the properties of projection inference for set-identified SVARs. 

A nominal $1-\alpha$ projection region collects all the structural parameters of interest that are compatible with the VAR reduced-form parameters in a nominal $1-\alpha$ Wald ellipsoid. By construction, projection inference does not rely on the specification of prior beliefs for set-identified parameters.

We showed that---under mild assumptions concerning the asymptotic behavior of estimators and posterior distributions for the reduced-form parameters--- projection produces regions with frequentist coverage and asymptotic \emph{robust} Bayesian credibility of at least $1-\alpha$.

The main drawback of our projection region is that it is conservative. For a frequentist, the large-sample coverage is strictly above the desired confidence level. For a robust Bayesian, the asymptotic robust credibility of the nominal $1-\alpha$ projection region is strictly above $1-\alpha$.  

We used the calibration idea described in \cite{Kaido_Molinari_Stoye:2014} to eliminate the excess of robust Bayesian credibility. The calibration procedure consists of drawing the reduced-form parameters, $\mu$, from its posterior distribution (or a suitable large-sample Gaussian approximation); evaluating the functions $\underline{v}(\mu), \overline{v}(\mu)$ for each draw of $\mu$; and, finally, decreasing the nominal level of the projection region until it contains exactly $(1-\alpha)\%$ of the values of $\underline{v}(\mu), \overline{v}(\mu)$. The calibration exercise required more work than the baseline projection, but it is computationally feasible (and easily parallelizable).  

We implemented our projection confidence set in a demand/supply SVAR of the U.S.~labor market. The main set-identifying assumptions were sign restrictions on contemporaneous responses. Standard Bayesian credible sets suggested that the medium-run response of wages and employment to structural shocks behave similar to the contemporaneous responses. Our projection region (baseline and calibrated) showed that only the qualitative effects of demand shocks on employment and the qualitative effects of supply shocks on wages are robust to the choice of prior. Our projection approach is a natural complement for the Bayesian credible sets that are commonly reported in applied macroeconomic work.


 \bibliography{revised_citations}

\appendix

\section*{Proof of  Theorem 5.1}  \label{subsection:R2}

The proof of Theorem \ref{result:R3} has already established that for any data realization:
$$\inf_{P^* \in \mathcal{P}(P^*_\mu)}P^* \Big( \lambda^{H}(A,B) \in CS_T (1-\alpha^* (Y_1, \ldots, Y_T); \lambda^H) \: \Big| \: Y_1, \ldots , Y_T \Big).$$
is at least as large as:
$$P^*_{\mu} \left( \times_{h=1}^{H} [\underline{v}_{k_h,i_h,j_h}(\mu), \overline{v}_{k_h,i_h,j_h}(\mu)] \subseteq \text{CS}_{T}(1-\alpha^*(Y_1, \ldots, Y_T), \lambda^{H})  |  Y_1, \ldots, Y_T  \right).$$  

Hence, it is sufficient to show that for any data realization:

$$\inf_{P^* \in \mathcal{P}(P^*_\mu)}P^* \Big( \lambda^{H}(A,B) \in CS_T (1-\alpha^*(Y_1, \ldots, Y_{T}); \lambda^H) \: \Big| \: Y_1, \ldots , Y_T \Big) \leq 1-\alpha.$$  

In order to establish this upper bound for each data realization, we will find a prior on $Q$ (conditional on $\mu$) that gives credibility of exactly $1-\alpha$ to the calibrated projection region. Fix the data, and denote the set $\text{CS}_T (1-\alpha(Y_1, \ldots, Y_{T}); \lambda^H)$ simply by $\mathcal{C}(Y^{T})$. Before the realization of the data, the set $\mathcal{C}(Y^{T})$ is just some subset of $\R^{H}$, so the prior can depend on this set. Let $\overline{v}_{h}(\mu)$ abbreviate $\overline{v}_{k_h,i_h,j_h}(\mu)$ and define $\underline{v}_{h}(\mu)$ analogously.  Let $Q_{\text{max}}(\mu; h)$ denote the rotation matrix for which the structural parameter achieves its upper bound; i.e., $\lambda(\mu, Q_{\text{max}}(\mu; h)) = \overline{v}_{h}(\mu)$ (the matrix $Q_{\min}$ is defined analogously). 

For each $\mu$ such that  $\times_{h=1}^{H} [\underline{v}_{h}(\mu), \overline{v}_{h}(\mu)] \notin \mathcal{C}(Y^{T})$, let $\overline{h}(\mu)$ denote the smallest horizon for which $\overline{v}_{\overline{h}(\mu)}(\mu)$ is not contained in the $h(\mu)$-th coordinate of the region $\mathcal{C}(Y^{T})$. If no upperbound falls outside $\mathcal{C}(Y^T)$ set $\overline{h}(\mu)=0$. Define $\underline{h}(\mu)$ analogously. Consider the following prior for $Q | \mu$ that depends on the set $\mathcal{C}_{T}(Y^{T})$: 
$$Q|\mu = 
\left\{
\begin{array}{ccc}
 Q_{\text{max}} (\mu; 1) & \text{if} & \times_{h=1}^{H} [\underline{v}_{h}(\mu), \overline{v}_{h}(\mu)] \subseteq \mathcal{C}_{T}(Y^T),     \\
 Q_{\text{max}} (\mu, \overline{h}(\mu)) & \text{if}  & \times_{h=1}^{H} [\underline{v}_{h}(\mu), \overline{v}_{h}(\mu)] \not\subseteq \mathcal{C}(Y^{T}) \text{ and }  \overline{h}(\mu) \geq \underline{h}(\mu),   \\
 Q_{\text{min}} (\mu, \underline{h}(\mu))  & \text{if}   &  \times_{h=1}^{H} [\underline{v}_{h}(\mu), \overline{v}_{h}(\mu)] \not\subseteq \mathcal{C} (Y^{T}) \text{ and }\overline{h}(\mu) < \underline{h}(\mu), 
\end{array}
\right.
 $$
Finally, let $P^{**}$ denote the prior induced by $P^*_{\mu}$ and $Q|\mu$ as defined above. Note that for each data realization $(Y_1, \ldots, Y_T):$
 $$\inf_{P^* \in \mathcal{P}(P^*_\mu)}P^* \Big( \lambda^{H}(A,B) \in CS_T (1-\alpha(Y_1, \ldots Y_{T}); \lambda^H) \: \Big| \: Y_1, \ldots , Y_T \Big)$$
 is---by definition of infimum---smaller than or equal 
$$ P^{**} \Big( \lambda^{H}(\mu,Q) \in CS_T (1-\alpha(Y_1, \ldots, Y_T); \lambda^H) \: \Big| \: Y_1, \ldots , Y_T \Big).$$
By construction, the prior for $Q|\mu$ is such that $\lambda^{H}(\mu,Q) \in CS_T (1-\alpha(Y_1, \ldots, Y_T); \lambda^H)$ if and only if $\times_{h=1}^{H} [\underline{v}_{h}(\mu), \overline{v}_{h}(\mu)] \subseteq \mathcal{C}_{T}(Y^{T})$. To see this, note that whenever the bounds of the identified set $\times_{h=1}^{H} [\underline{v}_{h}(\mu), \overline{v}_{h}(\mu)] \not\subseteq \mathcal{C}_{T}(Y^{T})$, either $\overline{h}(\mu) \neq 0$ or $\underline{h}(\mu) \neq 0$ implying that the structural parameter $\lambda_h(\mu, Q)$ takes the value of $\overline{v}_{\overline{h}(\mu)}(\mu)$ or $\underline{v}_{\overline{h}(\mu)}(\mu)$ (whichever horizon is largest). Since these bounds are not contained in $\mathcal{C}_{T}(Y^{T})$:
$$ P^{**} \Big( \lambda^{H}(\mu,Q) \in CS_T (1-\alpha(Y_1, \ldots, Y_T); \lambda^H) \: \Big| \: Y_1, \ldots , Y_T \Big).$$
equals 
$$P^*_{\mu} \left( \times_{h=1}^{H} [\underline{v}_{k_h,i_h,j_h}(\mu), \overline{v}_{k_h,i_h,j_h}(\mu)] \in \text{CS}_{T}(1-\alpha^*(Y_1, \ldots,Y_T), \lambda^{H})  |  Y_1, \ldots, Y_T  \right) = 1-\alpha.$$
This means that:
$$1-\alpha \leq \inf_{P^* \in \mathcal{P}(P^*_\mu)}P^* \Big( \lambda^{H}(A,B) \in CS_T (1-\alpha^*(Y_1, \ldots, Y_{T}); \lambda^H) \: \Big| \: Y_1, \ldots , Y_T \Big) \leq 1-\alpha.$$

\newpage
\renewcommand{\theequation}{S.\arabic{equation}}
\renewcommand{\thesection}{S\arabic{section}}
\renewcommand{\thepage}{S\arabic{page}}
\setcounter{equation}{0}
 
\setcounter{page}{1}
 



\section{Verification of Assumption 4.1 for the Gaussian SVAR with a Normal-Wishart Prior.} \label{subsection:NWprior}

Consider the SVAR in (3.1) and assume that $F \sim \mathcal{N}(0,\eye{n})$. Let $P^*$ denote a prior on the SVAR parameters $(A,B)$. 

Note first that Assumption 4.1 depends only on the distribution that $P^*$ induces over the reduced-form parameters, $\mu$. Thus, we abuse notation and refer to $P^*$ as the prior distribution on $(A,\Sigma)$. \\

The analysis in this section focuses on the \emph{Normal-Wishart} prior $P^*$ used in Gaussian SVAR analysis. We establish an almost sure version of Assumption 4.1. \\

{\scshape Prior for $\mu$:} Consider the hyper-parameters:
$$\bar{A}_0  \in  \R^{n \times np}, \: S_0 \in \R^{n \times n}, \: N_0 \in \R^{np \times np}, \: v_0 \in \R.$$

\begin{definition}
The Normal-Wishart Prior $P^*$ over the parameters $(\vect(A), \vech(\Sigma))$---defined by hyper-parameters $(\bar{A}_0, S_0, N_0, v_0)$---is given by:
$$ \vect(A) | \Sigma \sim \mathcal{N} \Big( \vect(\bar{A}_0) \: , \: N_0^{-1} \kron \Sigma \Big),$$
and
$$ \Sigma^{-1} \sim \text{Wishart}_n \Big( S_0^{-1} / v_0 \: , \: v_0 \Big).$$
\end{definition}
\noindent {\scshape Posterior in the Gaussian SVAR:} Let 
$$ Q_T \equiv \frac{1}{T} \sum_{t=1}^{T} X_t X_t',$$
and define the updated hyperparameters:

\begin{eqnarray*} \label{equation:PosteriorHyper}
\bar{A}_T &=& \widehat{A}_T Q_T \Big( \frac{N_0}{T} + Q_T \Big)^{-1} + \bar{A}_0 \frac{N_0}{T} \Big( \frac{N_0}{T} + Q_T \Big)^{-1} \\
S_T &=& \frac{v_0}{T+ v_0} S_0 + \frac{T}{T+v_0} \widehat{\Sigma}_T + \frac{1}{T+v_0} \Big(\bar{A}_T - \bar{A}_0\Big) N_0 \Big( \frac{N_0}{T} + Q_T \Big)^{-1} Q_T \Big(\bar{A}_T - \bar{A}_0\Big)'  \\
\end{eqnarray*} 
\noindent where $\widehat{A}_T$ and $\widehat{\Sigma}_T$ are the ordinary least squares estimators for $A$ and $\Sigma$ defined in Section 3.1. \\

From p. 410 in \cite{Uhlig:1994} and p. 410 in \cite{uhlig:2005}, the posterior distribution for the vector $(\vect(A)', \vech(\Sigma)')'$ can be written as:
$$ \vect(A) | Y_1, \ldots, Y_T  = \vect(\bar{A}_T) + \Big[  \Big( \frac{N_0}{T} + Q_T \Big)^{-1} \kron \frac{\Sigma}{T} \Big]^{1/2} W, \quad W\sim \mathcal{N}_{n^2p}(\bm{0}, \eye{n^2p}),$$
$$ \Sigma | Y_1, \ldots, Y_T  = S_T^{1/2} \Big( \frac{1}{T} \sum_{t=1}^{T} Z_t Z_t' \Big)^{-1} S_T^{1/2}, \quad Z_t \sim \mathcal{N}_{n}(\bm{0} \: , \: \eye{n}), \text{i.i.d},$$  
where both random vectors are independent of the data and $\{Z_t\}_{t=1}^{T}$ is independent of $W$. Note that for a given data realization, the posterior distribution of $(A,\Sigma)$ is a measurable function of $\mathcal{W} \equiv (W,Z_1, \ldots Z_T)$. We use the term $o_{\mathcal{W}}(1)$ to denote any sequence that converges to zero as $T \rightarrow \infty$ for almost every realization of $\mathcal{W}$.    \\

\noindent {\scshape Asymptotic Behavior of the posterior for $\mu$:} We now show that all of the Normal-Wishart priors in the Gaussian model satisfy our Assumption 4.1. Note first that for almost every data realization $(Y_1, \ldots , Y_T)$ and almost every realization of the random vector $Z_t$ we have that 
$$\Sigma - \widehat{\Sigma}_T \rightarrow 0,$$ 
by applying the strong law of large numbers to $(1/T) \sum_{t=1}^{T} Z_t Z_t'$. Consequently:\\
\begin{eqnarray*}
\sqrt{T} ( \vect(A)-\vect(\widehat{A}_T) ) &=&  \widehat{A}_T \sqrt{T} \Big( Q_T \Big( \frac{N_0}{T} + Q_T \Big)^{-1} - \eye{n^2 p} \Big) +\bar{A}_0 \frac{N_0}{\sqrt{T}} \Big( \frac{N_0}{T} + Q_T \Big)^{-1}  \\
&+& \Big[  \Big( \frac{N_0}{T} + Q_T \Big)^{-1} \kron \widehat{\Sigma}_{T} \Big]^{1/2} W + o_{P^* | Y_1, \ldots Y_T} (1),\\
&=& \widehat{A}_T \sqrt{T} \Big( Q_T \Big( Q_T^{-1} - Q_T^{-1}\frac{N_0}{T} Q_{T}^{-1} + O(1/T^2) \Big) - \eye{n^2 p } \Big) \\
&& \text{(by a first-order Taylor expansion)} \\
&+& \bar{A}_0 \frac{N_0}{\sqrt{T}} \Big( \frac{N_0}{T} + Q_T \Big)^{-1} \\
&+& \Big[  \Big( \frac{N_0}{T} + Q_T \Big)^{-1} \kron \widehat{\Sigma}_{T} \Big]^{1/2} W + o_{\mathcal{W}} (1), \\
\end{eqnarray*}
\begin{eqnarray*}
&=& \Big[   Q_T ^{-1} \kron \widehat{\Sigma}_{T} \Big]^{1/2} W + o_{\mathcal{W}} (1).\\
\end{eqnarray*}

This implies that the posterior distribution of $\sqrt{T} ( \vect(A)-\vect(\widehat{A}_T) )$ converges in distribution, for almost every data realization $(Y_1, \ldots , Y_T)$, to the random vector:

\begin{equation} \label{equation:G1}
 [Q_T ^{-1/2} \kron \widehat{\Sigma}_{T}^{1/2}] W, \text{ where } W\sim \mathcal{N}_{n^2p}(\bm{0}, \eye{n^2p}).
\end{equation}

\noindent Note now that 

\begin{eqnarray*}
\sqrt{T} ( \vech(\Sigma)-\vech(\widehat{\Sigma}_T) ) &=&  \sqrt{T} \vech \Big(S_T^{1/2} \Big( \frac{1}{T} \sum_{t=1}^{T} Z_t Z_t' \Big)^{-1} S_T^{1/2} - \widehat{\Sigma}_{T} \Big), \\ 
&=&  \sqrt{T} \vech \Big(\widehat{\Sigma}_T^{1/2} \Big( \frac{1}{T} \sum_{t=1}^{T} Z_t Z_t' \Big)^{-1} \widehat{\Sigma}_T^{1/2} + O(1/T) - \widehat{\Sigma}_{T} \Big), \\
&=&  \sqrt{T} \vech\Big( \widehat{\Sigma}^{1/2}_T \Big[ \Big(\frac{1}{T} \sum_{t=1}^{T} Z_t Z_t' \Big)^{-1} - \eye{n} \Big] \widehat{\Sigma}_T^{1/2} \Big)+ o(1). \\
\end{eqnarray*}

This implies that the posterior distribution of $\sqrt{T} ( \vech(\Sigma)-\vech(\widehat{\Sigma}_T) )$ converges in distribution, for almost every data realization $(Y_1, \ldots , Y_T)$, to the random vector:

\begin{equation}\label{equation:G2}
\Big( 2 D^{+} (\widehat{\Sigma}_{T} \kron \widehat{\Sigma}_{T} ) (D^{+})^\prime  \Big)^{1/2} Z, \text{ where } Z \sim \mathcal{N}_{n(n+1)/2} (\textbf{0}, \eye{n(n+1)/2}), \: Z \bot W, \\ 
\end{equation}

\noindent and $D^{+} \equiv (D'D)^{-1}D'$ is the Moore-Penrose inverse of the duplication matrix D such that $\vect(\Sigma)= D \vech(\Sigma)$. 

Now, assume that the confidence set for the reduced-form parameters is constructed using the Gaussian Maximum Likelihood asymptotic variance of $\widehat{\mu}_{T}$ as in p.93 of \cite{Lutkepohl:2007}; that is:

\begin{equation}\label{equation:G3}
\widehat{\Omega}_{T} \equiv \begin{pmatrix} Q_T^{-1}  \kron \widehat{\Sigma}_T & \mathbf{0}_{n^2 p \times (n(n+1)/2)} \\ \mathbf{0}_{(n(n+1)/2) \times n^2 p} &  2 D^{+} (\widehat{\Sigma}_{T} \kron \widehat{\Sigma}_{T} ) {D^{+}}'   \end{pmatrix}. 
\end{equation}

Let $G$ denote the joint distribution of $(W,Z)$, which is a standard multivariate normal independently of the data. Then, combining (\ref{equation:G1}), (\ref{equation:G2}), (\ref{equation:G3}) 

\begin{eqnarray*}
P^* \Big(  \mu \in \text{CS}_T(1-\alpha, \mu)  | (Y_1, \ldots , Y_T) \Big) &=& P^* \Big(  \sqrt{T}(\mu-\widehat{\mu}_{T})' \widehat{\Omega}_T^{-1} \sqrt{T}(\mu- \widehat{\mu}_{T}) \leq \chi^2_{d,1-\alpha}  | (Y_1, \ldots Y_T) \Big) \\
& \rightarrow & G \Big(  \begin{pmatrix} W \\ Z \end{pmatrix} ' \begin{pmatrix} W \\ Z \end{pmatrix} \leq  \chi^2_{d,1-\alpha} \: | \: Y_1, \ldots , Y_T \Big) \text{for a.e. data realization} \\
&=& G \Big(  \begin{pmatrix} W \\ Z \end{pmatrix} ' \begin{pmatrix} W \\ Z \end{pmatrix} \leq  \chi^2_{d,1-\alpha} \Big)  \\
&=& (1-\alpha). 
\end{eqnarray*}

\section{Asymptotic Calibration for a Robust Bayesian. } \label{subsection:AsyCali}

We now show that whenever $\alpha^*_{T} \equiv \alpha(Y_1, \ldots, Y_{T})$ is calibrated to guarantee that
$$ \prob_{T} \Big(  \times_{h=1}^{H} [\underline{v}_{k_1,i_1,j_1}(\mu), \overline{v}_{k_1,i_1,j_1}(\mu)] \times \ldots \times [\underline{v}_{k_h,i_h,j_h}(\mu), \overline{v}_{k_h,i_h,j_h}(\mu)] \subseteq \text{CS}_{T}(1-\alpha^*_{T}, \lambda^{H}) \: | \: Y_1, \ldots Y_{T} \Big)$$
equals $1-\alpha$ whenever $\mu | Y_1, \ldots Y_{T} \sim \mathcal{N}_{d}(\widehat{\mu}_{T},\widehat{\Omega}_{T}/T)$, then one can guarantee asymptotic robust credibility of $1-\alpha$ for a large class of priors on $\mu$. This is formalized below. 

Let $f(Y_1, \ldots Y_{T} \: | \: \mu_0)$ denote the Gaussian density for the VAR data and let $\Omega \in \R^{d \times d}$ denote the probability limit of $\widehat{\Omega}_{T}$. Let $G_{\Omega}$ denote a Gaussian measure centered at $\mathbf{0}_{d}$ with covariance matrix $\Omega$. Let $\mathcal{B}(d)$ denote Borel sets in $\R^{d}$.

\begin{lemma}
Let $Y_1, \ldots Y_{T} \sim f(Y_1, \ldots Y_{T} \: | \: \mu_0)$ and suppose that the prior $P^*_\mu$ is such that:
$$ \sup_{A \in \mathcal{B}(d)} \left | P^*_{\mu} ( \sqrt{T}(\mu - \widehat{\mu}_{T}) \in A \: | \: Y_1, \ldots Y_{T} ) -G_{\Omega}(A ) \right | = o_{p}(Y_1, \ldots Y_{T}; \mu_0).$$

\noindent Then, 
$$ \inf_{P^* \in \mathcal{P}(P^*_{\mu})} P^* \Big(  \lambda^{H}(A,B) \in  \text{CS}_{T}(1-\alpha^*_{T}, \lambda^{H}) \: | \: Y_1, \ldots Y_{T} \Big)=1-\alpha + o_p(Y_1, \ldots Y_{T}; \mu_0).$$

\end{lemma}

\begin{proof}
Theorem 5.1 has shown that for any $\alpha(Y_1, \ldots ,Y_{Y})$
$$ \inf_{P^* \in \mathcal{P}(P^*_{\mu})} P^* \Big(  \lambda^{H}(A,B) \in  \text{CS}_{T}(1-\alpha^*_{T}, \lambda^{H}) \: | \: Y_1, \ldots Y_{T} \Big) = P^*_{\mu} \left( \mu \in A^*_{T} \: | \: Y_1, \ldots Y_{T} \right),$$
where $A^*_{T} \subseteq \R^{d}$ is defined as:
$$\{\mu \in \R^{d} \: | \: \times_{h=1}^{H} [\underline{v}_{k_h, i_h, j_h}(\mu),\overline{v}_{k_h, i_h, j_h}(\mu) ] \subseteq  \text{CS}_{T}(1-\alpha^*_{T}, \lambda^{H}) \}.$$
Note that
\begin{eqnarray*}
P^*_{\mu} \left( \mu \in A^*_{T} \: | \: Y_1, \ldots Y_{T} \right) &=& P^*_{\mu} \left( \sqrt{T}(\mu - \widehat{\mu}_{T})  \in \sqrt{T}(A^{*}_{T}-\widehat{\mu}_{T}) \ : | \: Y_1, \ldots Y_{T} \right)-G_{\Omega}(\sqrt{T}(A^{*}_{T}-\widehat{\mu}_{T})) \\
&+& G_{\Omega}(\sqrt{T}(A^{*}_{T}-\widehat{\mu}_{T})) - G_{\widehat{\Omega}_{T}}(\sqrt{T}(A^{*}_{T}-\widehat{\mu}_{T}))\\
&+ & G_{\widehat{\Omega}_{T}}(\sqrt{T}(A^{*}_{T}-\widehat{\mu}_{T}))  
\end{eqnarray*}
We make three observations:\\
\begin{enumerate}
\item Note first that:
$$ P^*_{\mu} \left( \sqrt{T}(\mu - \widehat{\mu}_{T})  \in \sqrt{T}(A^{*}_{T}-\widehat{\mu}_{T}) \: | \: Y_1, \ldots Y_{T} \right)-G_{\Omega}(\sqrt{T}(A^{*}_{T}-\widehat{\mu}_{T}))  $$
is smaller than or equal  
$$ \sup_{A \in \mathcal{B}(d)} \left | P^*_{\mu} ( \sqrt{T}(\mu - \widehat{\mu}_{T}) \in A \: | \: Y_1, \ldots Y_{T} ) -G_{\Omega}(A ) \right |,$$
which is, by assumption, $o_{p}(Y_1, \ldots Y_{T}; \mu_0).$ \\
\item Note then that 
$$|G_{\widehat{\Omega}_{T}}(\sqrt{T}(A^{*}_{T}-\widehat{\mu}_{T})) - G_{\Omega}(\sqrt{T}(A^{*}_{T}-\widehat{\mu}_T))| = o_{p}(Y_1, \ldots, Y_{T}; \mu_0)$$
since $\widehat{\Omega}_{T} \cprob \Omega$ and $G$ is the Gaussian measure centered at zero.
\item Finally, note that $G_{\widehat{\Omega}_{T}}(\sqrt{T}(A^{*}_{T}-\widehat{\mu}_{T}))$ is the same as

$$\prob ( N(\widehat{\mu}_{T}, \widehat{\Omega}_{T}/T)  \in A^*_{T} | Y_1, \ldots, Y_{T} ),$$
which, by definition of $A^*_{T}$, is the same as:
$$ \prob_{T} \Big(  \times_{h=1}^{H} [\underline{v}_{k_1,i_1,j_1}(\mu), \overline{v}_{k_1,i_1,j_1}(\mu)] \times \ldots \times [\underline{v}_{k_h,i_h,j_h}(\mu), \overline{v}_{k_h,i_h,j_h}(\mu)] \subseteq \text{CS}_{T}(1-\alpha^*_{T}, \lambda^{H}) \: | \: Y_1, \ldots Y_{T} \Big) $$
where $\mu | Y_1, \ldots Y_{T} \sim \mathcal{N}_{d}(\widehat{\mu}_{T},\widehat{\Omega}_{T}/T)$. 
\end{enumerate}
We conclude that:
$$ |\inf_{P^* \in \mathcal{P}(P^*_{\mu})} P^* \Big(  \lambda^{H}(A,B) \in  \text{CS}_{T}(1-\alpha^*_{T}, \lambda^{H}) \: | \: Y_1, \ldots Y_{T} \Big) - (1-\alpha) |  \leq o_p(Y_1, \ldots Y_{T}; \mu_0),$$
which implies the desired result.  
\end{proof}

\section{Addenda for implementation} 
\label{sec:implementation}


{\scshape The nature of the optimization problem:} The nonlinear mathematical program in~(6.1) has two challenging features. On the one hand, the optimization problem is non-convex; this complicates the task of finding a global minimum with algorithms designed to detect local optima. On the other hand, the number of optimization arguments and constraints increases quadratically in the dimension of the SVAR; this compromises the feasibility of some optimization routines designed to detect global optima (for example, brute-force grid search on CS$_T(1-\alpha, \mu)$ to optimize $\underline{v}_{k,i,j}(\mu)$ and $\overline{v}_{k,i,j}(\mu)$).  

{\scshape Our Approach:} Taking these two features into consideration, we first implemented projection by running a local optimization algorithm followed by a global algorithm that used the local solution as an input. The algorithms and the functions used to implement the projection confidence interval are described below. In the application analyzed in this paper, the global stage of the algorithm did not have any impact on the local solution. We thus suggest researchers to implement our approach using only the SQP/IP routine described below. 

{\scshape Local Algorithms:} Although no standard classification exists for local optimization algorithms, the most common procedures are often grouped as follows: penalty and Augmented Lagrangian Methods; Sequential Quadratic Programming (SQP); and Interior Point Methods (IP); see p. 422 of \cite{Nocedal:06} for more details.

Within this class of algorithms, we focus on the IP and SQP algorithms, both of which are considered as the ``\textit{most powerful algorithms for large-scale nonlinear programming}'', \cite{Nocedal:06}, p. 563.\footnote{Furthermore, these algorithms exploit the existence of second-order derivatives which are well-defined in our problem.} Conveniently, IP and SQP are included in Matlab\textsuperscript{\textregistered}'s \texttt{fmincon} function, which comes with the Optimization toolbox. We run the SQP algorithm---which is usually faster than IP---and in case it does not find a solution, we switch to IP, an algorithm which we denote by \emph{SQP/IP}.

{\scshape Global Algorithms:} IP and SQP are well suited to handle various degeneracy problems in order to find a local minimum for large-scale non-convex problems.  There is now a large body of literature on global optimization strategies; see \cite{horst:1995} and \cite{Pardalos:2013}.  Popular global optimization algorithms include adaptive stochastic search; branch and bound methods; homotopy methods; Genetic algorithms (GA); simulated annealing and two-phase algorithms such as \emph{MultiStart} and \emph{GlobalSearch}.\footnote{For a more detailed list and classification of global methods see p. 519 of Chapter 15 in \cite{Pardalos:2013}. For a description of two-phase algorithms see Chapter 12 in \cite{Pardalos:2013}. }We focus on the two-phase algorithms MultiStart, GlobalSearch and on the genetic algorithm available in Matlab.\footnote{Genetic algorithms are a well-developed field of computing and they have been used in many applications; see the introduction to Chapter 9 in \cite{Pardalos:2013}. A very interesting application in economics that motivated our focus on GA is given in \cite{Qu:2015}.}

\begin{figure}[H] 
\centering
\caption{Accuracy of SQP/IP for a demand shock}
{\includegraphics[keepaspectratio, scale=.45]{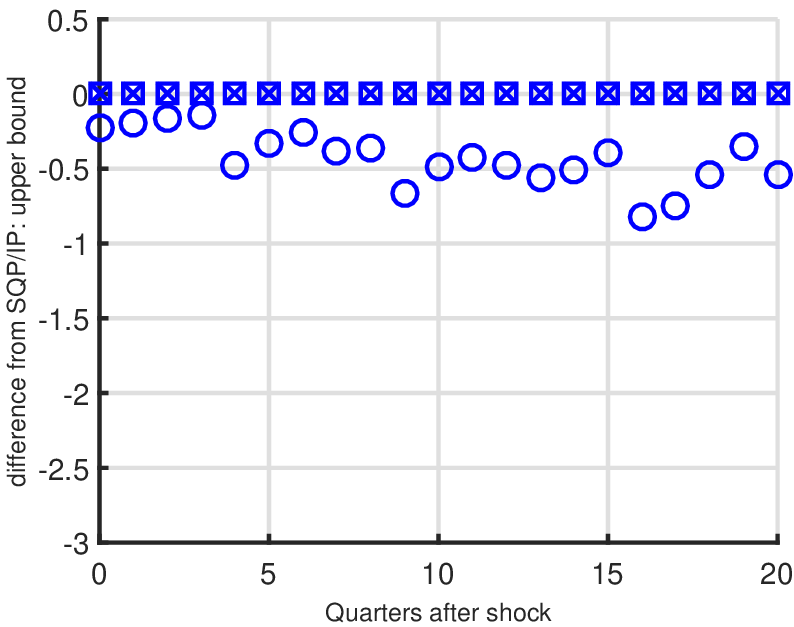}}\hspace{.2cm}
{\includegraphics[keepaspectratio, scale=.45]{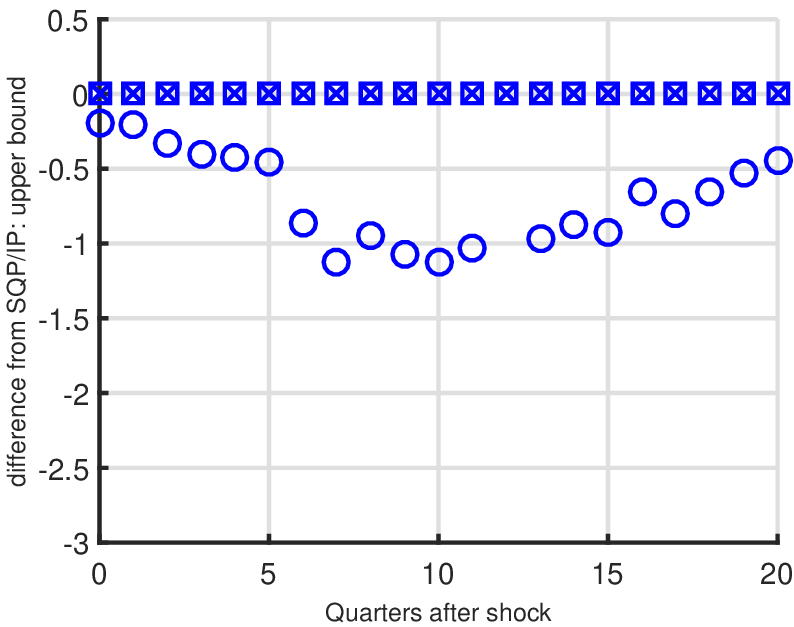}}\\
\subfloat[Wage Response]{\includegraphics[keepaspectratio, scale=.45]{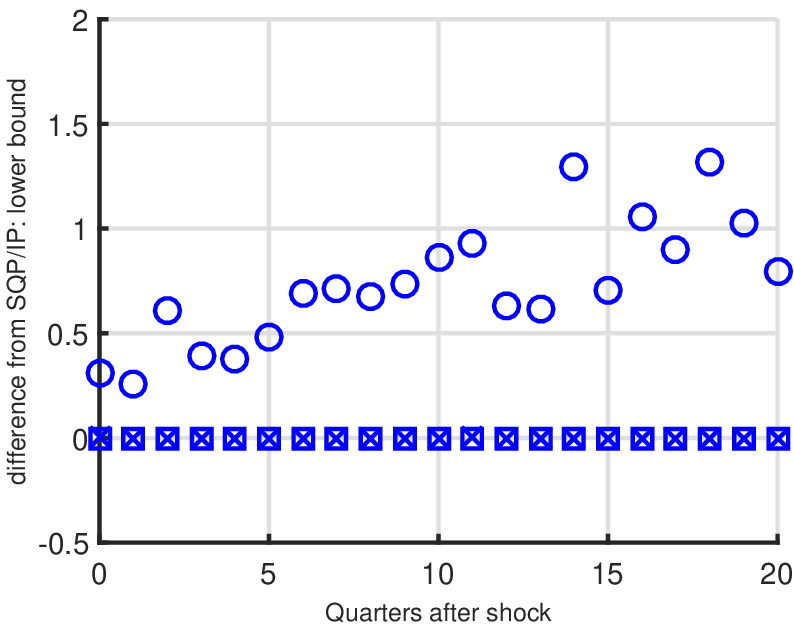}}\hspace{.2cm}
\subfloat[Employment Response]{\includegraphics[keepaspectratio, scale=.45]{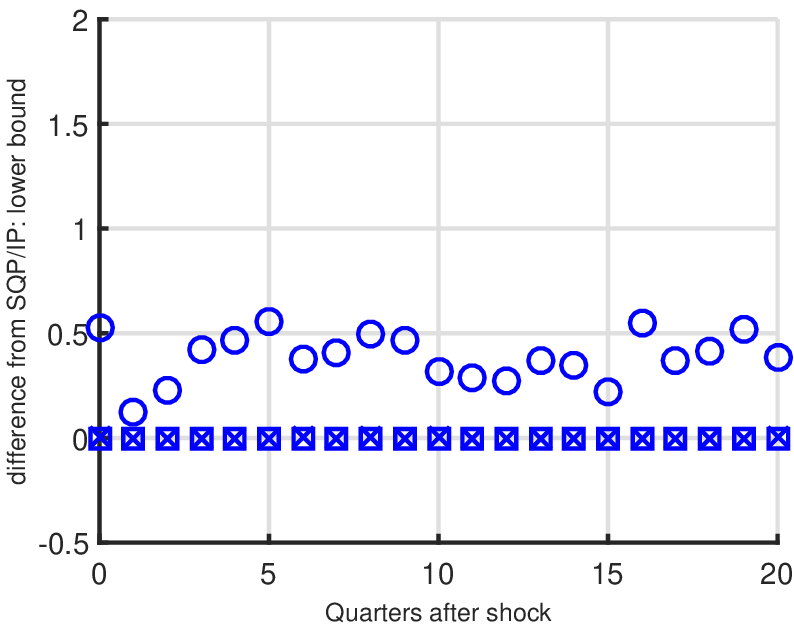}}\\
\caption*{\footnotesize ({\scshape Square, Blue}) Optimal Value reported by SQP/IP minus Optimal Value reported by SQP/IP + \texttt{Multistart};  ({\scshape Cross, Blue}) Optimal Value reported by SQP/IP minus Optimal Value reported by SQP/IP + \texttt{Global Search}; ({\scshape Circle, Blue}) Optimal Value reported by SQP/IP minus Optimal Value reported by \texttt{GA}. All vertical axes are expressed in percentage points.}
\label{figure:accuracy}
\end{figure}

{\scshape Comments Regarding Local and Global algorithms:} Figure~\ref{figure:accuracy} compares the bounds of the projection confidence interval for the first four algorithms listed in Table~\ref{table:T3}. For this application, it seems that none of the global algorithms improve on the local solution obtained from SQP/IP.

\begin{table}[H] \caption{Computational time in seconds}\label{table:time_cost}
\begin{center} 
\begin{tabular}{l l r r r}
\hline\hline \vspace{-1mm}\\
Algorithm & Details & Time \\
\hline\\
SQP/IP			 		&& 734 \\
\vspace{-3mm}\\
SQP/IP + MultiStart		& 100 initial points & 33,314 \\
SQP/IP + GlobalSearch		& 100 trial points (20 in Stage 1) & 1,359 \\
Genetic Algorithm		& population of 100, 500 generations & 76,863 \\
\vspace{-3mm}\\
Grid Search on CS$_{T}(1-\alpha,\mu)$ 			& 1,000 draws from $\mu$ & 4,548 \\
\vspace{-3mm}\\
Bayesian, BH			& 1,000,000 Metropolis-Hastings draws  & 3,992 \\
Bayesian, Uhlig 	& 100,000 accepted posterior draws & 2,338 \\
\vspace{-2mm}\\ \hline\hline \vspace{-2mm}\\
\multicolumn{3}{l}{Notes: Laptop @2.4GHz IntelCore i7.}
\end{tabular} \label{table:T3}
\end{center} \end{table}

\newpage

\section{Additional Figures} 

\newpage

\begin{figure}[H] 
\centering
\caption{Simulation error in Projection region.}
{\includegraphics[keepaspectratio, scale=.45]{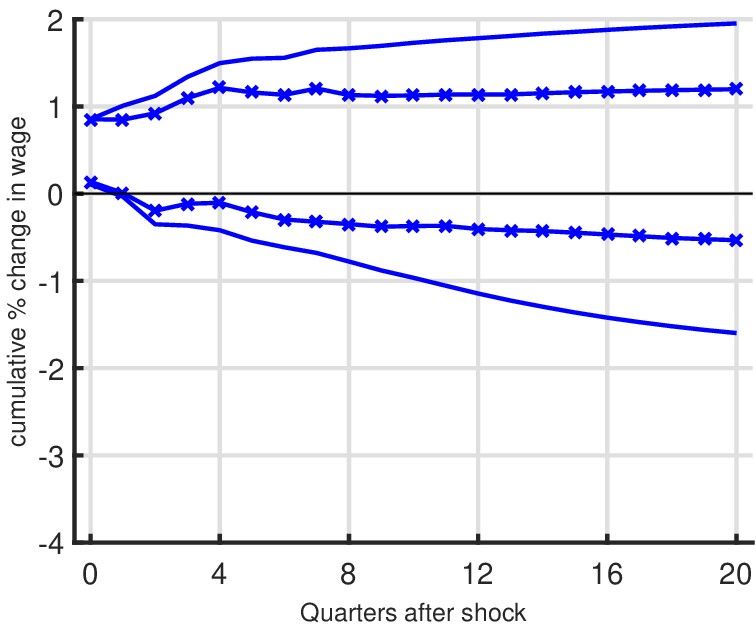}} \hspace{.5cm}
{\includegraphics[keepaspectratio, scale=.45]{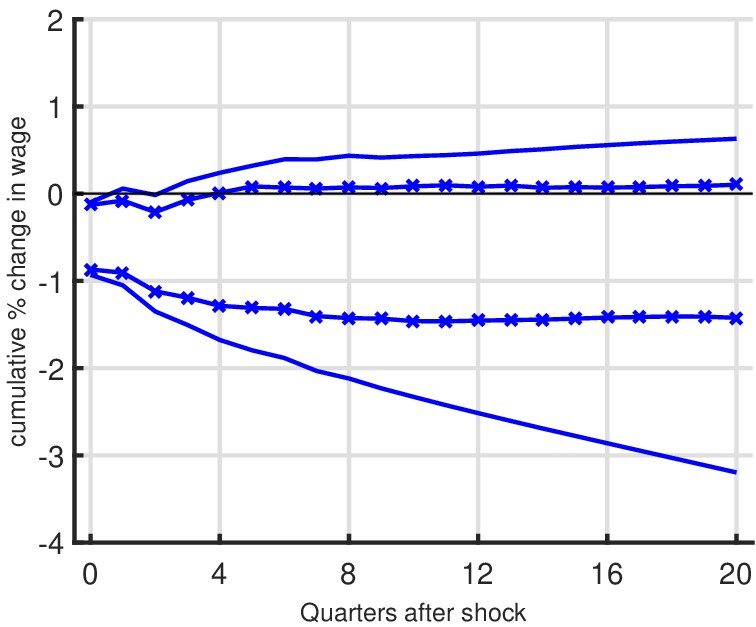}}\\
\subfloat[Expansionary Demand Shock]{\includegraphics[keepaspectratio, scale=.45]{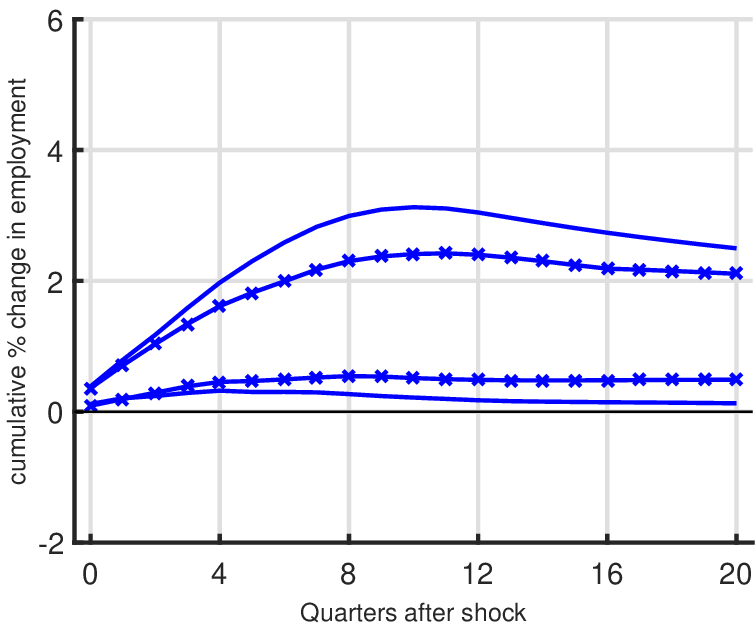}} \hspace{.5cm}
\subfloat[Expansionary Supply Shock]{\includegraphics[keepaspectratio, scale=.45]{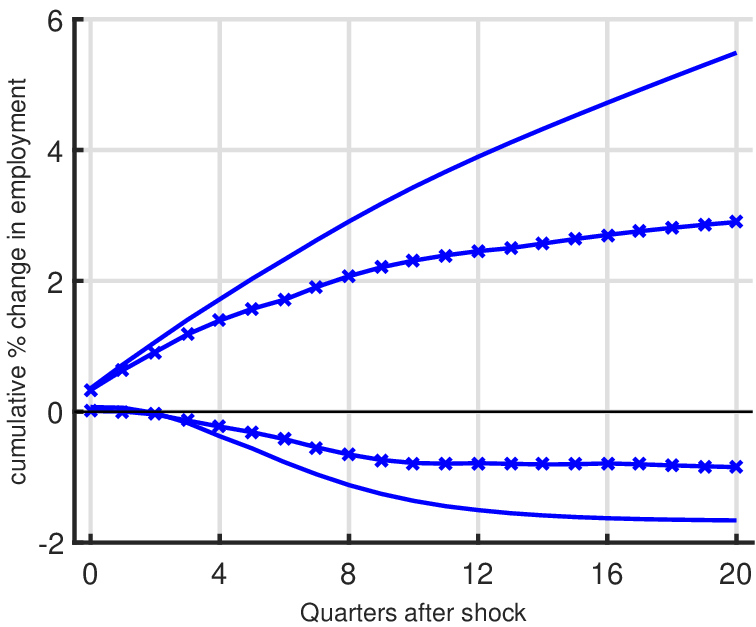}}\\
\label{figure:simerror}
\caption*{\footnotesize ({\scshape Solid Line}) 68\% Projection region using the SQP/IP algorithm described in Section \ref{sec:implementation}; ({\scshape Connected, Solid Line}) 68\% Projection region using a two-step algorithm: 1) Sample M=100,000 reduced form parameters that satisfy the 68\% Wald ellipsoid constraint. 2) For each draw, solve for the identified set. The smallest and largest value of the identified set is the simulation-based approximation of the Projection region.}
\end{figure} 



\begin{figure}  [H]
\centering
\caption{68\% Projection Region and 68\% Credible Set.}
\caption*{(\cite{uhlig:2005} priors)} 
{\includegraphics[keepaspectratio, scale=.45]{./figures/figImplementation/IRF_v1_s1_p6.eps}} \hspace{.5cm}
{\includegraphics[keepaspectratio, scale=.45]{./figures/figImplementation/IRF_v1_s2_p6.eps}}\\
\subfloat[Expansionary Demand Shock]{\includegraphics[keepaspectratio, scale=.45]{figures/figImplementation/IRF_v2_s1_p6.eps}} \hspace{.5cm}
\subfloat[Expansionary Supply Shock]{\includegraphics[keepaspectratio, scale=.45]{./figures/figImplementation/IRF_v2_s2_p6.eps}}\\
\label{figure:Uhlig}
\caption*{\footnotesize ({\scshape Solid, Blue Line}) 68\% Projection region;  ({\scshape Shaded, Gray Area}) 68\% Bayesian Credible Set based on the Normal-Wishart-Haar priors suggested in \cite{uhlig:2005} and the inequality constraints summarized below Table~2. The credible set is implemented following  \cite{Arias:2015}. }
\end{figure}

\end{document}